\documentclass[fleqn,usenatbib]{mnras}

\usepackage{newtxtext,newtxmath}

\usepackage[T1]{fontenc}

\DeclareRobustCommand{\VAN}[3]{#2}
\let\VANthebibliography\thebibliography
\def\thebibliography{\DeclareRobustCommand{\VAN}[3]{##3}\VANthebibliography}

\usepackage{graphicx}	
\usepackage{amsmath}	
\usepackage{siunitx}
\usepackage{xcolor}

\usepackage{orcidlink}  

\title[The dynamical origin of BCGs and ICL]{Disentangling the assembly of the central and diffuse stellar components in galaxy clusters through dynamical friction and tidal stripping}

\author[F. R. Pearce et al.]{
Frazer R. Pearce\orcidlink{0000-0002-2383-9250}$^{1}$\thanks{E-mail: frazer.pearce@nottingham.ac.uk},
Steven P. Bamford\orcidlink{0000-0001-7821-7195}$^{1}$,
Garreth Martin\orcidlink{0000-0003-2939-8668}$^{1}$,
Yannick M. Bah\'{e}\orcidlink{0000-0002-3196-5126}$^{1,2}$,
\newauthor~Harley J. Brown\orcidlink{0009-0004-7521-8204}$^{1}$,
Madeline Butler\orcidlink{0009-0004-9273-4630}$^{1}$,
Ana Contreras-Santos\orcidlink{0000-0002-3374-4626}$^{3,4,5}$,
Weiguang Cui\orcidlink{0000-0002-2113-4863}$^{5,6,7}$,
\newauthor~Nina Hatch\orcidlink{0000-0001-5600-0534}$^{1}$,
and Alexander Knebe\orcidlink{0000-0003-4066-8307}$^{5,6,8}$
\\
$^{1}$University of Nottingham, School of Physics and Astronomy, Nottingham NG7 2RD, UK\\
$^{2}$Laboratoire d'Astrophysique, \'{E}cole Polytechnique F\'{e}d\'{e}rale de Lausanne (EPFL), Observatoire de Sauverny, 1290 Versoix, Switzerland\\
$^{3}$Departamento de Astrofísica, Universidad de La Laguna, Av. Astrofísico Francisco Sanchez s/n, 38206 La Laguna, Tenerife, Spain\\
$^{4}$Instituto de Astrofísica de Canarias, C/Vía Láctea s/n, 38205 La Laguna, Tenerife, Spain\\
$^{5}$Departamento de Física Te\'orica, M\'odulo 15, Facultad de Ciencias, Universidad Aut\'onoma de Madrid, 28049 Madrid, Spain\\
$^{6}$Centro de Investigaci\'on Avanzada en F\'isica Fundamental (CIAFF), Facultad de Ciencias, Universidad Aut\'onoma de Madrid, 28049 Madrid, Spain\\
$^{7}$Institute for Astronomy, Royal Observatory, Edinburgh EH9 3HJ, UK\\
$^{8}$International Centre for Radio Astronomy Research, University of Western Australia, 35 Stirling Highway, Crawley, Western Australia 6009, Australia
}

\date{Accepted XXX. Received YYY; in original form ZZZ}

\pubyear{\the\year{}}

\begin{document}
\label{firstpage}
\pagerange{\pageref{firstpage}--\pageref{lastpage}}
\maketitle

\begin{abstract}
In galaxy clusters, it is common to conceptually distinguish the central, brightest cluster galaxy (BCG) from the more diffuse intracluster light (ICL). So far, however, this distinction has lacked a clear physical basis, confusing both theoretical and observational studies. We address this by presenting a dynamically motivated definition of the two stellar components, evaluated using cluster zoom simulations from \textsc{the Three Hundred} project. We start with a motivational hindcasting approach, robustly tracing stars from the centres of 107 present-day clusters back through the merger hierarchy. We find that the majority of the final BCG stellar mass is typically contributed by the central galaxies of only 3--5 progenitor haloes. Most accreted galaxies contribute no stars to the BCG, instead surviving as satellites or depositing stellar material throughout the cluster. This naturally leads to a definition of the BCG stellar population as originating from the central galaxies of the most massive infalling systems. These are delivered to the cluster centre via efficient dynamical friction and subsequently undergo violent relaxation. The ICL arises from the complementary population of stars tidally stripped from the outskirts of all infalling systems. We therefore develop a forecasting method, identifying stars in the BCG and ICL via the dynamical fates of their originating stellar systems. Applying this to 261 clusters, we find the resulting components remain spatially interleaved, but occupy distinct regions of phase space. Our physically defined BCG matches conventional aperture-based selections with high completeness, while providing a direct interpretation in terms of hierarchical assembly.
\end{abstract}




\section{Introduction}

The stellar mass within a galaxy cluster is conventionally divided into three components: the brightest cluster galaxy (BCG) at the cluster centre, smaller orbiting satellite galaxies, and the intracluster light (ICL). In practice, however, the distinction between the BCG and the ICL is not clean. Observationally, the outer envelope of the BCG blends smoothly into diffuse low-surface-brightness light extending to large radii. In simulations, once stars bound to surviving subhaloes are removed, the remaining stellar distribution associated with the main halo is similarly continuous \citep{Gonzalez2005,Zibetti2005,Montes2018, Montes2022}. A wide range of operational definitions have therefore been adopted to separate the BCG from the ICL. In simulations, common choices include fixed physical apertures around the central galaxy, often of order $30$--$100\, \mathrm{kpc}$ \citep[e.g.][]{Pillepich2018,ContrerasSantos2024} or a multiple of the stellar half-mass radius \citep[e.g.][]{Ahvazi2024, Montenegro2025}. In observations, the ICL is frequently identified using isophotal thresholds \citep[e.g.][]{Zibetti2005,Furnell2021}. Dynamical approaches instead use unbinding criteria \citep[e.g.][]{Dolag2010}, or split the stellar component kinematically into a dynamically cold central component embedded within a hotter diffuse stellar envelope \citep[e.g.][]{Cui2014, Longobardi2015, Remus2017}. More generally, projected phase-space information can also be used to distinguish halo components \citep[e.g.][]{Longobardi2015,Marini2024}. Reviews and broader comparisons of these approaches are given by \citet{Contini2021} and \citet{Brough2024}. No single choice has emerged as uniquely physical or universally applicable and several authors have chosen to avoid making the distinction altogether.

This ambiguity is not merely a naming convention. As emphasized by \citet{Pillepich2018}, no optimal generalized physical transition between the bright inner galaxy and its lower surface-brightness envelope is readily identifiable. The light profile of the central galaxy often merges smoothly into the ICL, making an unambiguous definition of the stellar mass of the central galaxy difficult. They accordingly advocate fixed apertures as the most practical and least ambiguous operational definition. They also show that, in the most massive haloes, the diffuse stellar mass outside 100~kpc can exceed that within 100~kpc, and that the outer stellar density profile can be as shallow as that of the dark matter. This is a useful practical viewpoint, but it also underlines the central problem: the BCG--ICL split is usually imposed on a continuous stellar distribution rather than read directly from a sharply bimodal one.

One might expect the BCG and ICL to arise from different assembly channels, with mergers building the BCG and stripping building the ICL.  Hierarchical models have long shown that present-day BCGs are assembled from many progenitors over extended periods of time, with a large fraction of their stars formed ex situ and only later brought together through repeated accretion events \citep{DeLucia2007}. Massive galaxies are often interpreted within a two-phase formation picture, in which an early phase of rapid in-situ assembly is followed by later growth through the accretion of ex-situ stellar material, particularly in their outer regions \citep{Oser2010}. Minor mergers are also widely thought to contribute significantly to the structural growth of massive galaxies through inside-out growth of their outer envelopes, and to the late assembly of BCG outskirts \citep{Naab2009,Hilz2013,Laporte2013,Burke2015}. Conversely, recent work on ICL assembly suggests that the dominant progenitors of ICL stars are typically not tiny dwarfs, but rather satellites with infall stellar masses above ${\sim} 10^{10}\,{\rm M_\odot}$, with the main contribution often arising from galaxies around $10^{10.5}$--$10^{11.5}\,{\rm M_\odot}$ \citep{Contini2014,Contini2019,Brown2024,HBrown2026, Martin2026}.

A further complication is that even the inferred amount of ICL depends strongly on how the BCG--ICL separation is implemented. In a recent homogeneous comparison across four hydrodynamical simulation suites, \citet{HBrown2026} showed that while the absolute ICL fraction is highly sensitive to the adopted definition, the inter-simulation differences become much smaller once a common methodology is imposed. They also showed that whether the average ICL fraction appears to evolve strongly with redshift or remain roughly constant is itself largely driven by the specific BCG--ICL definition adopted. Indeed, because of this arbitrariness, several studies now analyse the combined BCG+ICL system rather than insisting on a hard decomposition \citep{Pillepich2018,HBrown2026}.

Rather than deciding where an arbitrary boundary should be drawn, it is more useful to ask which stars should most naturally be regarded as belonging to the central galaxy. In a simulated cluster this can be explored more readily than in observations. Stars bound to surviving subhaloes belong to satellite systems. Among the remaining stars associated with the dark matter halo of the cluster itself, some contribute to the central galaxy, while the rest populate a diffuse stellar component orbiting in the cluster potential. In that sense, if one could physically extract the BCG from the cluster halo, the residual diffuse stellar component would automatically be the ICL. This does not imply a clean physical separation in configuration or phase space: the two populations may remain intertwined. Rather, it suggests that the primary problem is to identify the central-galaxy stars in a physically motivated way.

Recent dynamical work has indicated that a phase-space perspective can be informative. \citet{Butler2025} showed that intracluster stars are not simply a generic tracer population following the dark matter: they have lower specific energies, more centrally concentrated density profiles, and more radially biased orbits than the dark matter. In addition, these conclusions are robust to the precise BCG--ICL separation adopted. Thus, even though the morphology of the diffuse stellar component may broadly track the host halo, its occupation of phase space is biased. By extension, stars associated with the long-lived central galaxy should occupy an even more tightly bound and lower-dispersion subset of the cluster stellar distribution than the ICL. The issue is therefore not whether one can find a perfectly clean separator in radius or velocity space. The BCG and ICL may be physically distinct yet spatially overlapping populations. Instead, the challenge is to define the central-galaxy component through its assembly history and dynamical fate, and then ask whether this physically motivated decomposition suggests a useful observational boundary.

The novelty of this paper is that we define BCG membership by assembly history and dynamical fate, rather than by present-day radius, binding energy, surface brightness, or kinematics. Specifically, we build the BCG from the central stellar material of progenitors that are subsequently dynamically incorporated into the main halo and treat the remaining non-satellite cluster stars as the diffuse intracluster component. The resulting BCG--ICL decomposition is therefore not imposed as a geometric cut at $z=0$, but derived from the merger history of the cluster and the fate of accreted stellar systems.

This perspective motivates two complementary approaches. The first is retrospective, or hindcasting. We select stars associated with the $z=0$ central region of a cluster using a conventional practical definition, such as a fixed aperture, and trace them backwards through the assembly history. This determines where the stars formed and which progenitors brought them in. Hindcasting asks which objects actually contributed stars to the present-day central region of the cluster. The second is predictive, or forecasting: given an accretion event, determine whether its stars are incorporated into a structure at the centre of the cluster, or instead remain part of the extended cluster population. This forecasting approach reframes the problem away from arbitrary apertures and towards the competition between two physical pathways. Dynamical friction delivers infalling objects to the cluster centre, where violent relaxation incorporates them into the existing central structure. Tidal stripping and disruption instead disperse their stars at larger radii \citep{Murante2007,Contini2014,Contini2018,Contini2021}.

In this paper we use these two approaches together. We begin with the traditional hindcasting approach, then mirror this with a forecasting construction. This yields a physically selected BCG stellar component and a complementary ICL population. We then examine the radial and kinematic distributions of these two populations. The radius at which their contributions become comparable is interpreted not as a sharp edge, since no such edge is expected, but as a characteristic BCG scale derived from assembly physics rather than imposed \emph{a priori}.

Our aim is therefore not to argue that BCG and ICL are fundamentally distinct stellar components separated by a unique structural boundary. Rather, it is to show that the continuous stellar distribution of the cluster halo can be decomposed in a physically interpretable way by identifying the stars that belong to a central structure formed through violent relaxation. This provides a bridge between conventional post hoc definitions of the BCG and a forward model based on the dynamical fate of accreted stellar systems, and may in turn suggest more robust observational proxies for separating central-galaxy light from the diffuse intracluster component.

This paper proceeds as follows: in Section~\ref{sec:sims} we introduce the simulations used, our choice of halo analysis, and summarise relevant physical processes. In Section~\ref{sec:hindcast} we explore the hindcasting approach of tracing stellar populations selected at $z=0$ backwards in time to see where they came from. In Section~\ref{sec:forecast} we reverse this and implement forecasting, asking what is the eventual fate of infalling galaxies, developing a model of which events build the central BCG. In Section~\ref{sec:implications} we discuss the implications of this model and in Section~\ref{sec:conclusions} we conclude by summarising our key findings.

\section{Simulations, halo catalogues, merger trees and relevant physical processes}
\label{sec:sims}

Our analysis primarily uses the \textsc{GIZMO\_3k} cluster zoom simulations from \textsc{The Three Hundred} project \citep{Cui2018,Cui2022}, except for a convergence test that utilises a subset of 45 existing higher resolution \textsc{GIZMO\_7k} counterparts of these runs. \textsc{The Three Hundred} comprises 324 hydrodynamical re-simulations of the most massive clusters drawn from the MultiDark-Planck2 parent simulation \citep{klypin_MD}, with each zoom region centred on a cluster-scale halo and extending to a radius of $15\,h^{-1}\,\mathrm{Mpc}$ from the cluster centre. The runs analysed here were evolved with the meshless finite-mass hydrodynamics solver \textsc{GIZMO} \citep{Hopkins2015}, using a \textsc{SIMBA}-based galaxy-formation model \citep{Dave2019}. The \textsc{SIMBA-C} form \citep{Hough2023,Hough2024} was used for the \textsc{GIZMO\_7k} runs. These have been retuned to account for the increased resolution and upgraded subgrid physics. The adopted cosmology is that of the MultiDark-Planck simulation, with
$(\Omega_{\rm m},\Omega_{\Lambda},\Omega_{\rm b},\sigma_8,n_{\rm s},h)
= (0.307,0.693,0.04825,0.8288,0.9611,0.6777)$.
In the high-resolution region, the particle masses are approximately
$m_{\rm DM}\simeq 1.3\times10^9\,{\rm M_\odot}$ and
$m_{\rm gas}\simeq 2.4\times10^8\,{\rm M_\odot}$ (eight times smaller for the \textsc{GIZMO\_7k} runs). Within these simulations, haloes and subhaloes are identified with the Amiga Halo Finder, \textsc{AHF} \citep{Knollmann2009}. Throughout this paper, $R_{200}$ denotes the radius enclosing a mean density 200 times the critical density, with $M_{200c}$ the corresponding enclosed mass. Where fixed physical apertures originate from quantities expressed in $h^{-1}$ units, we give both forms when first introduced. Radii expressed as $r/R_{200}$ refer to the final $z=0$ host halo unless explicitly identified as progenitor radii.

To connect structures between snapshots, we use the merger trees produced by the \textsc{MergerTree} algorithm distributed with \textsc{AHF}. In particular, we use the \texttt{skipping+stars} variant, which patches over temporary tree breaks and uses both dark matter and stellar particles in the merit function, as first described in \citet{ContrerasSantos2024}. This choice is helpful in dense cluster environments, where stripping, close passages, and transient mis-identifications can otherwise produce broken or erratic ancestry links. The use of particle-overlap-based tree building follows standard practice in merger-tree construction and comparison studies \citep{Srisawat2013}. \textsc{AHF} provides particle membership and particle type lists for all the structures it finds. By default \textsc{AHF} produces {\it inclusive} particle membership lists, i.e. a particle is listed as being a member of all the structures it is bound to. In this work we solely use {\it exclusive} membership lists, where a particle is only listed as a member of the deepest level of the structure hierarchy. i.e. in this work a particle belongs solely to the least massive structure it is bound to. This exclusive assignment is important for the present analysis, because it prevents stars associated with resolved substructure from being counted simultaneously as members of both a satellite and the diffuse stellar component of its host. The veracity of the \textsc{MergerTree} scheme is evidenced by the fact that all but four of the 324 cluster merger-trees can be traced back successfully to beyond $z=3.5$. 

We follow a similar but bespoke merit function based methodology to trace the main progenitor of stars occupying the central region to earlier times. Specifically we take the particle IDs of all the stars within $147.6\,\mathrm{kpc}\ ( 100\,h^{-1}\,\mathrm{kpc})$ of the cluster centre at $z=0$ and find the least massive halo each of them belongs to at the previous snapshot. We then choose the halo owning the largest number of matching stars as the progenitor. This search is recursively repeated at earlier snapshots using the matched list of stellar particle IDs. This methodology also produces long continuous histories for the stars occupying the central region at $z = 0$. In all 324 clusters we can trace the main progenitor of the central stars back to beyond $z=3.5$.  

\subsection{Checking consistency between halo and galaxy main branches}
\label{sec:tracking_clean}

A practical difficulty in any particle-based assembly analysis is that the object being tracked is not unique in a major merger. There are, in effect, two coupled but distinct entities: the dark-matter halo, identified by the halo finder and merger tree, and the galaxy at its centre, identified by its stellar content. In an idealized case, the cluster halo tree and the evolutionary sequence of the central galaxy would remain aligned. In practice, they need not do so.

The first issue is the familiar ambiguity of halo identification during major mergers. Once two comparably massive haloes overlap significantly, the partition of particles between them can become unstable from one snapshot to the next, and the labels of ``main halo'' and ``subhalo'' may swap even when the underlying physical system evolves smoothly. This behaviour has been documented in halo-finder and merger-tree comparison studies, particularly for close major mergers where the two systems are difficult to separate robustly
\citep{Srisawat2013, Avila2014, BehrooziMajor2015, Han2018}. In such cases the nominal halo main branch can become unreliable, or terminate prematurely. In practice, \textsc{MergerTree} constructs the cluster halo main branch back further than 12\,Gyr for all except four of our 324 clusters, with all main branches extending back to lookback times above 10\,Gyr ($z = 1.77$).

A second issue is that even if the cluster main branch is formally preserved, the galaxy associated to the cluster centre may change. During a close encounter between two massive systems, the dominant stellar core may switch from one galaxy halo to another. Essentially, there can be several large galaxies near the centre of the cluster halo and the structure finder has to choose which one is dominant. For the present problem this matters because the hindcasting analysis relies on continuity of stellar identity. If the galaxy halo associated with the cluster changes then the stellar mass associated with the centre of the cluster can jump  because a large number of stars suddenly seem to appear.

We therefore test, for each cluster, whether the cluster main branch and the galaxy associated with it remain consistent. Formally, we compare the cluster main branch and the main progenitor of the stars in the central region snapshot-by-snapshot and check that they agree. 
The 107 clusters for which the cluster main branch and the central star main branch remain identical to lookback times greater than \(10\,\mathrm{Gyr}\) (\(z=1.77\)) are retained for the hindcasting analysis below. This is a deliberately conservative requirement, but it ensures that the subsequent particle tracking is performed only in systems where the central halo and the central galaxy have a consistent ancestry over the period relevant to the present study. Prior to this epoch there are a plethora of similar possible histories that could be followed, all containing relatively small galaxies compared to the final object. Which of these histories is chosen has no impact on the results presented here. Note that cluster 1, the largest cluster in the study, is one of the clusters that fails this test, swapping central galaxy briefly at $z = 0.59$. For this reason we use cluster 2, the largest cluster that satisfies this test, as the example cluster in the plots below.

\subsection{Dynamical friction}
\label{subsec:dynfric}

For a satellite of mass \(M_{\rm sat}\) orbiting within a host of mass \(M_{\rm host}\), the timescale for orbital decay increases rapidly as the mass ratio becomes more extreme. In the theoretical picture, dynamical friction is stronger for more massive perturbers because the deceleration scales approximately with the perturber mass \citep{Chandrasekhar1943}. In realistic cosmological mergers the dependence is more complicated, but the same basic conclusion remains: high mass-ratio mergers take much longer to sink, while nearly equal mass-ratio mergers decay much more quickly \citep{vanDenBosch1999,BoylanKolchin2008,Amorisco2017}.

One merger-time estimate is the fitting formula of \citet{BoylanKolchin2008},
\begin{equation}
t_{\rm merge} = A\,
\frac{(M_{\rm host}/M_{\rm sat})^{b}}
{\ln\!\left(1+M_{\rm host}/M_{\rm sat}\right)}
\exp(c\eta)
\left(\frac{r_{\rm c}(E)}{r_{\rm vir}}\right)^d
t_{\rm dyn},
\label{eq:bk08}
\end{equation}
where \(\eta\) is the orbital circularity, \(r_{\rm c}(E)\) is the circular-orbit radius of the same orbital energy, and \(t_{\rm dyn}\) is the host dynamical time. We take the standard coefficients ($A  = 0.216, b = 1.3, c = 1.9, d = 1.0)$ from \citet{BoylanKolchin2008}. Even allowing for the spread in orbital parameters seen in cosmological infall \citep{Wetzel2011, Martin2026}, the expected merger time rises steeply as \(M_{\rm sat}/M_{\rm host}\) decreases. Massive satellites therefore have a plausible route to becoming part of the central galaxy within a Hubble time; low-mass satellites generally do not. 

\subsection{Tidal stripping}
\label{subsec:stripping}

While dynamical friction acts to bring satellites inwards, tidal stripping acts to remove their mass, largely during their first central passage. As a satellite orbits through the host potential, its outer dark-matter halo is stripped first, reducing the bound mass that drives further dynamical friction. Continued stripping can then remove stars, especially from loosely bound outer stellar envelopes \citep{Read2006,Smith2016,Contini2018,he2026}. Essentially, tidal stripping acts to lengthen the dynamical friction timescale of infalling objects \citep{Hashimoto_2003, Gan2010, Tamfal_2021}.

The stripped stellar material does not subsequently behave like an intact satellite core. Once unbound, those stars join the diffuse cluster potential and no longer experience the coherent orbital decay of a self-bound galaxy. They may later contribute to the central regions statistically, but not through the direct delivery of an infalling central stellar system. By contrast, the dense inner cores of sufficiently massive galaxies can survive long enough to reach the centre and merge violently with the existing BCG. The stars carried by those cores are therefore far more likely to become part of the central galaxy itself.

The consequence is a strong mass-ratio dependence in the fate of accreted stellar mass. Massive progenitors are more likely to deposit their central galaxy directly into the cluster centre; lower-mass progenitors are more likely to contribute stars to the diffuse intracluster component, and/or remain bound in long-lived satellites.

\subsection{Violent relaxation}
\label{subsec:violent_relaxation}

In a strongly radial, near head-on encounter -- including those following orbital decay due to dynamical friction -- the gravitational potential changes rapidly on a timescale comparable to the orbital time of particles in the inner regions. This leads to violent relaxation, in which particle energies are redistributed as the system settles towards a new equilibrium \citep{LyndenBell1967}. This is not merely a numerical nuisance: the process phase-mixes stars formed in disparate objects to form a distinctly new object. Violent mergers have also been discussed as a physical channel for liberating stars from merging galaxies and contributing to the diffuse intracluster component \citep{Murante2007,Dolag2010,Contini2021}.

A concern in any particle-tracking analysis is whether violent relaxation during a major merger can compromise the integrity of particle ancestry. We tested the most extreme radial encounters in our sample, namely cases where an infalling system passes directly through the cluster centre. For these events we examined whether there was any evidence for a wholesale exchange of particle membership between the central halo and the infaller. We find no sign of such behaviour. Although the encounters do produce the expected strong mixing associated with violent relaxation, the particle IDs associated with each progenitor remain traceable through the merger, and there is no tendency for the two systems simply to exchange their stellar contents.

This is an important practical result. It implies that, while violent relaxation may contribute physically to the generation of diffuse stellar material, the ICL identified in this work is not being produced spuriously through numerical particle swapping in head-on mergers. The diffuse stellar component recovered by our analysis is therefore associated with genuine stripping, incomplete central assimilation, and hierarchical assembly, rather than with an artefact of the particle bookkeeping.

\section{Tracing central galaxy stars backwards}
\label{sec:hindcast}

\begin{figure*}
\includegraphics[width=16.5cm]{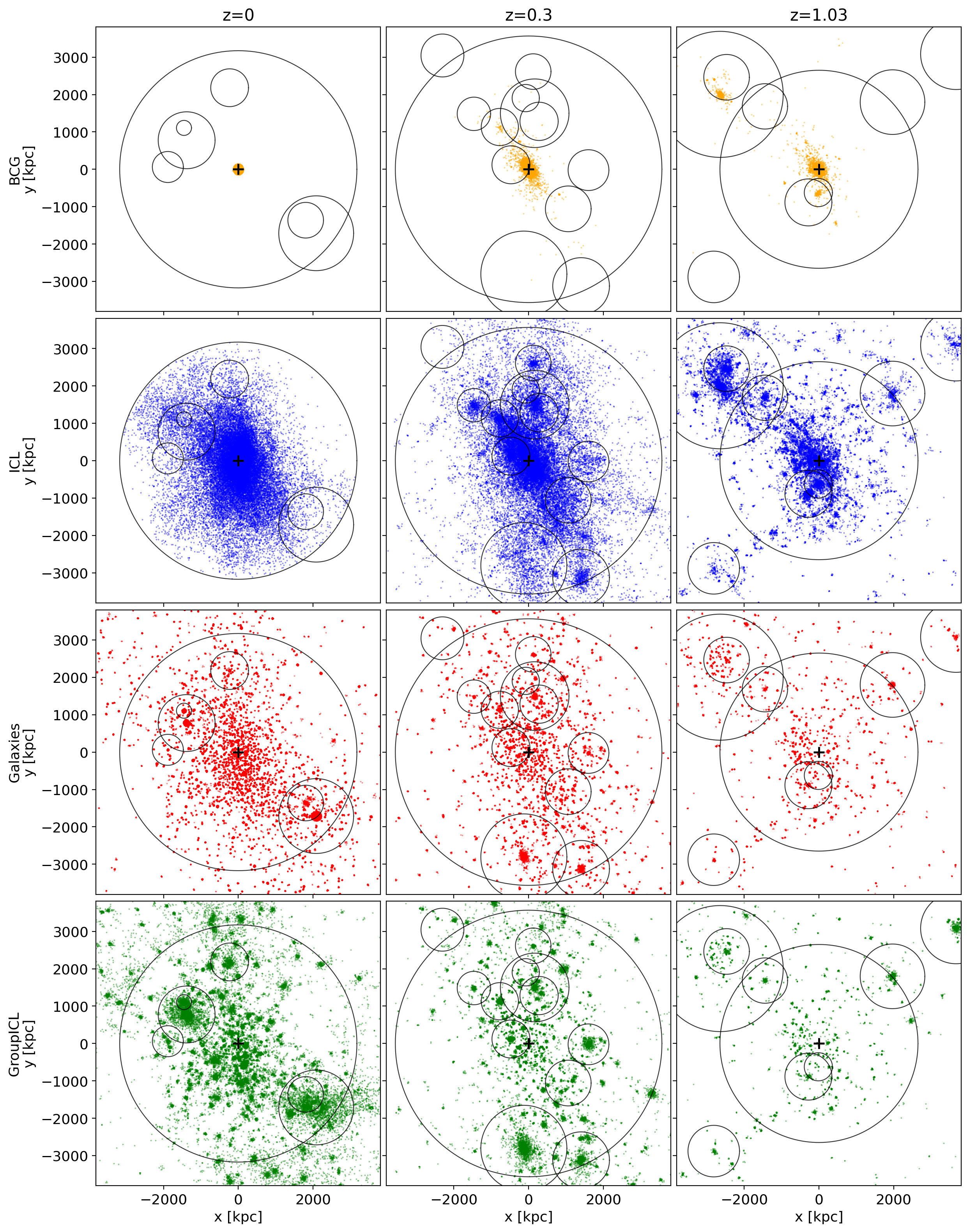}
    \caption{
Hindcast decomposition of the stellar population of cluster 2, the second largest object in the MDPL2 volume with $M_{200} = 3.4 \times 10^{15}\, \mathrm{M}_\odot$. Left column, $z=0$, middle column $z=0.3$, right column $z \sim 1$. All star particles within a cube of side length $2.4R_{200}$, centred on the cluster, are divided into four components at $z=0$. The top row shows the BCG, defined as stars bound to the cluster and lying within 147.6~kpc of the halo centre. The second row shows the cluster ICL, comprising the remaining stars associated with the cluster. The third row shows stars assigned to satellite galaxies, identified as those lying within $\min(2r_{1/2,\star},147.6\,{\rm kpc})$ of each non-central halo. The bottom row shows the remaining non-central halo stars, labelled GroupICL.}   
\label{fig:hindcast_split_z0}
\end{figure*}

\begin{figure}
	\includegraphics[width=\columnwidth]{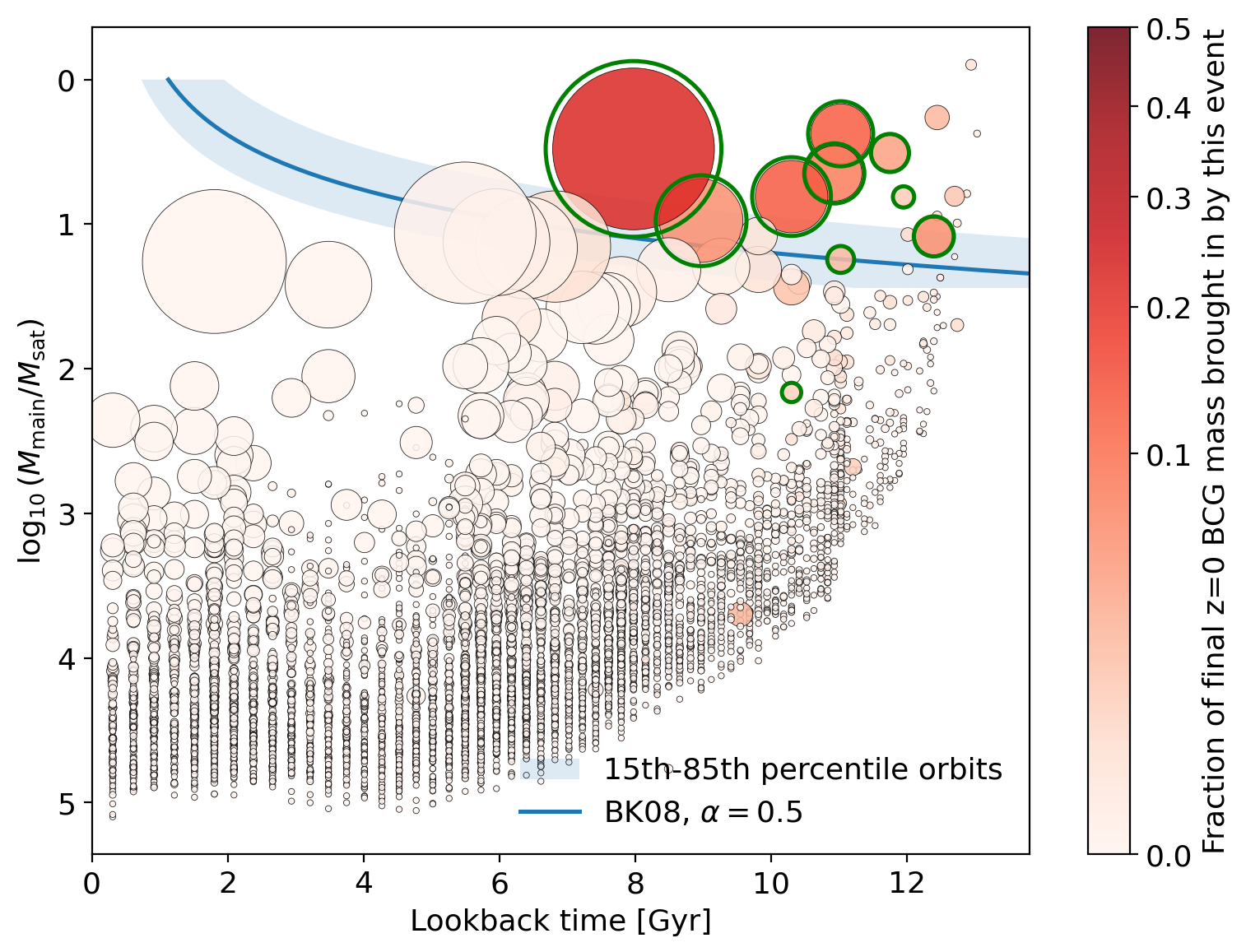}
    \caption{All captured halo infall events contributing to the assembly of cluster 2 as a function of merger lookback time and infall mass ratio. Each point corresponds to an object crossing $R_{200}$ of the cluster at that time. The vertical axis shows $\log_{10}(M_{\rm main}/M_{\rm sat})$, where the mass ratio is defined using the prior maximum masses of the cluster and the infalling system.  The symbol area is proportional to the total incoming plus post-infall stellar mass associated with that infall event, while the symbol colour indicates the fraction of the final stellar mass within the BCG aperture contributed by that event. The solid curve shows the \citet{BoylanKolchin2008} dynamical friction merger timescale, the shaded region indicates the corresponding orbital scatter range. The green
    rings indicate all the infall events contributing to the BCG component of the forecasting model detailed in Section~\ref{sec:forecast} below.}
\label{fig:cluster_buildup}
\end{figure}

\begin{figure}	\includegraphics[width=\columnwidth]{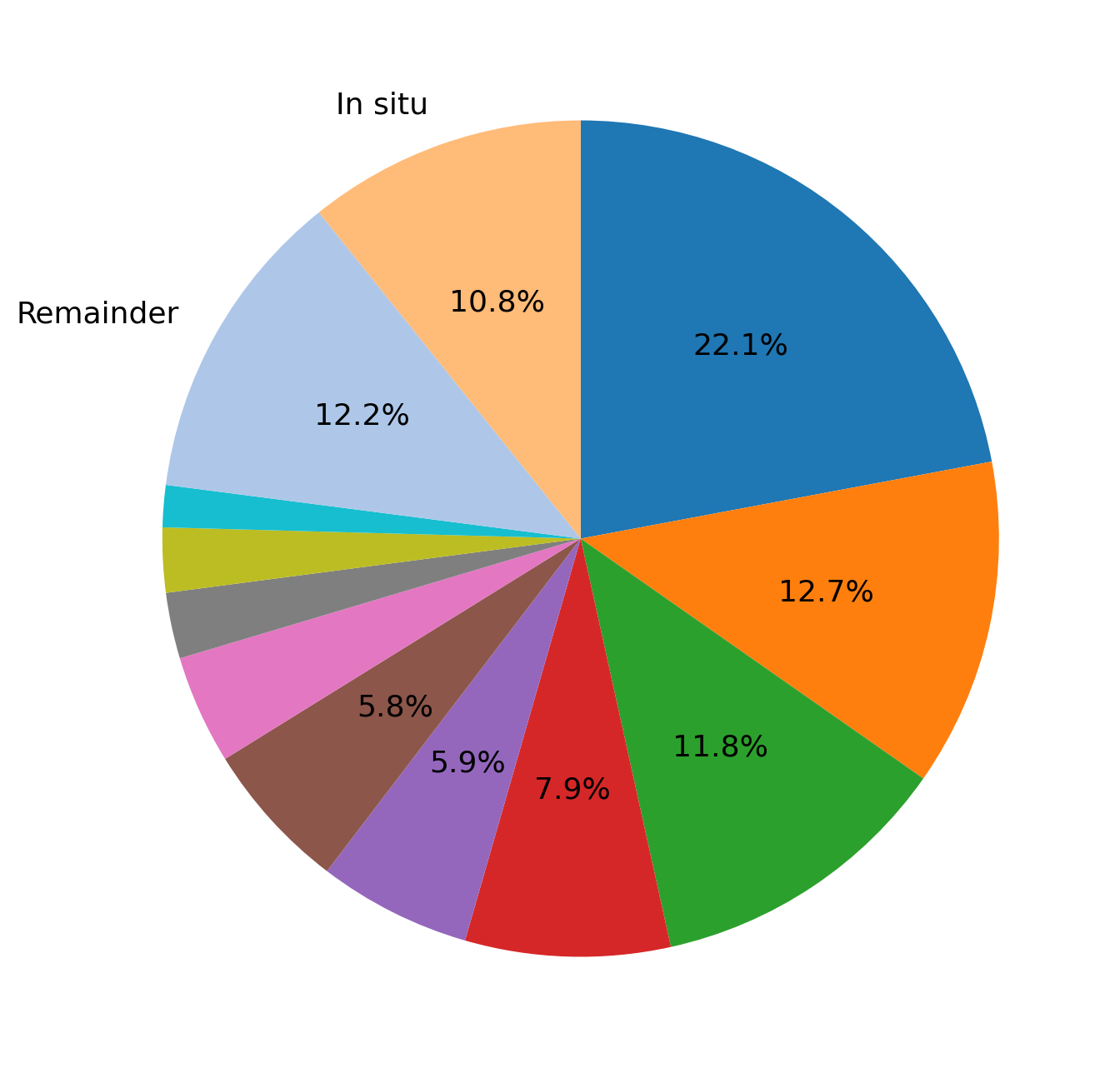}
\caption{BCG aperture stellar buildup: each of the first 10 slices show the ranked mass fraction of the final BCG aperture stellar mass associated with a single infall event identified within the cluster assembly history given in the event distribution of Figure~\ref{fig:cluster_buildup} for cluster 2. Infall events contributing more than $5\%$ of the final BCG aperture stellar mass are labelled individually. The final two slices show the remainder fraction, the amount of stars brought in by all other mergers and the in-situ fraction of stars forming within the cluster halo (see text).} 
\label{fig:bcg_buildup_pie}
\end{figure}

\begin{figure}
	\includegraphics[width=\columnwidth]{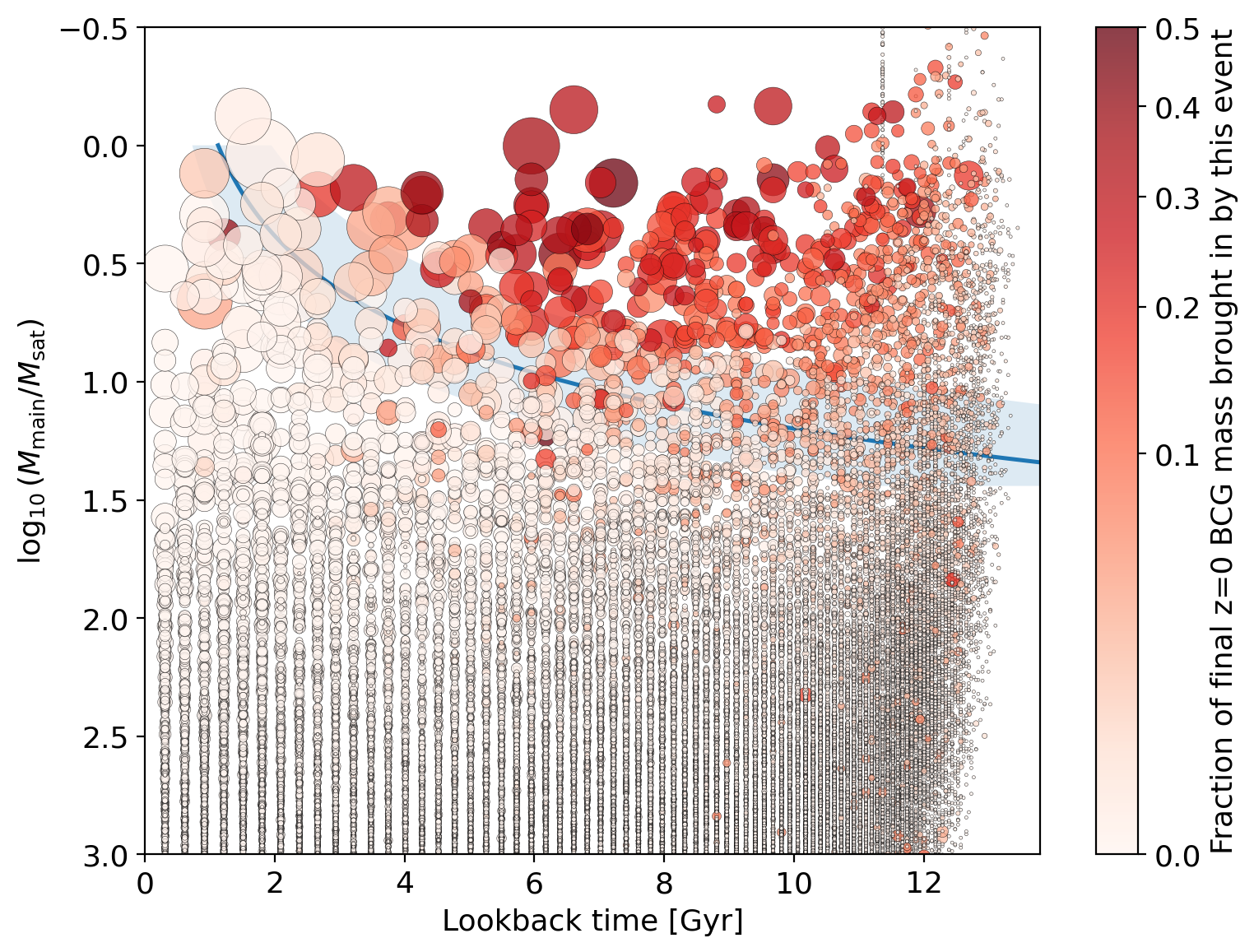}
\caption{As for Figure~\ref{fig:cluster_buildup}, but stacked for the 107 clusters with perfectly matching cluster and BCG histories back to $10\, \mathrm{Gyr} (z = 1.77)$. The plotted mass ratio range is truncated compared to Figure~\ref{fig:cluster_buildup} because very few events with mass ratios above 1000 contribute to the central BCG aperture. Areas are scaled down by a factor of 5  relative to those used in Figure~\ref{fig:cluster_buildup} for clarity. The solid blue curve shows the \citet{BoylanKolchin2008} dynamical friction merger timescale, the shaded region indicates the corresponding orbital scatter range. The plumes of small haloes with small merger mass ratios to the top right of the figure are examples of what happens when the cluster and galaxy tracking fails to follow the same object. Alignment is only guaranteed to a lookback time of $10\,\mathrm{Gyr}$. } 
\label{fig:bcg_average}
\end{figure}

\begin{figure}
	\includegraphics[width=\columnwidth]{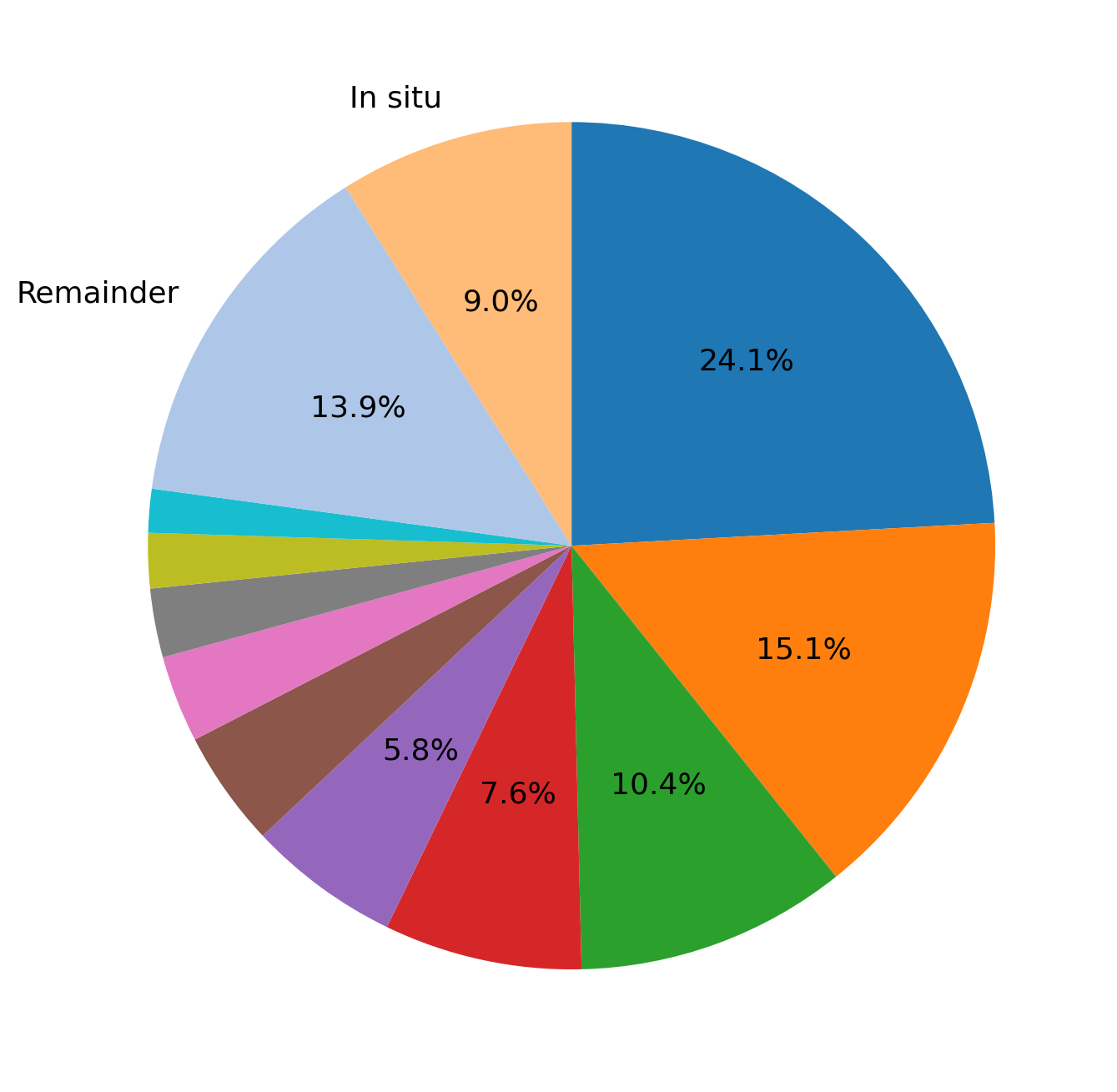}
\caption{As for Figure~\ref{fig:bcg_buildup_pie}, but averaged over the 107 clusters selected for Figure~\ref{fig:bcg_average}. For each cluster, the 10 events contributing the largest mass of stars to the BCG aperture are ranked in size, stacked across all clusters and averaged. The remaining infall events are combined into the remainder slice. Averaged in-situ star formation within the cluster centre is also indicated.}
\label{fig:bcg_average_pie}
\end{figure}

A straightforward way to study the origin of the present-day BCG and ICL is to start from the stars that are present near the centre at \(z=0\) and trace them backwards through the merger tree. This is, in some sense, an observationally motivated approach: one begins with the final stellar configuration, partitions it into components that resemble familiar observational categories, and then asks where those stars resided at earlier times. It is therefore easy to interpret, easy to visualise, and provides a direct complement to our physics-based forecasting analysis presented later. The price paid for this simplicity is that the method is intrinsically retrospective: it tells us where the final stars came from, but the selection at $z=0$ is arbitrarily imposed rather than derived.

We first select a target halo as the largest halo within $3\,\mathrm{Mpc}$ of the centre of each zoom region at $z=0$. We then divide the stars contained in this object into four categories:
(i) stars belonging to the central halo and lying within a fixed aperture of $100\,h^{-1}\,{\rm kpc}=147.6\,{\rm kpc}$ for our adopted value $h=0.6777$. This is intended as a conventional, deliberately broad BCG aperture, comparable to fixed apertures commonly adopted in simulation studies, rather than as a physically preferred BCG--ICL boundary. Its purpose here is simply to provide a well-defined $z=0$ stellar population whose assembly history can be traced backwards. We label this population ``BCG'';
(ii) stars belonging to the central halo but lying outside $147.6\,\mathrm{kpc}$, labelled ``ICL'';
(iii) stars associated with other galaxies in the region, labelled ``Galaxies''; and
(iv) the remaining diffuse stellar component around those non-central galaxies, labelled ``GroupICL''. The galaxy component is identified using the stellar half-mass radii;  \(r_{\ast,1/2}\), computed from the enclosed stellar mass profile. Stars within
\begin{equation}
r_{\rm gal}=\min\!\left(2\,r_{\ast,1/2},\,147.6\,\mathrm{kpc}\right)
\end{equation}
of that halo centre are classified as belonging to the galaxy, with the remainder assigned to the diffuse group-scale ICL component.  This division is used to generate the four-panel maps in Figure~\ref{fig:hindcast_split_z0}. 

The middle and right columns of Figure~\ref{fig:hindcast_split_z0} show the locations of the \(z=0\) stellar particle divisions mapped back to \(z=0.30\) and \(z=1.03\) and centred on the location of the main progenitor of the $z=0$ cluster at this time. The remapping is done by identifying the main progenitor halo from the \textsc{AHF} merger tree and then locating those particle IDs that exist in the earlier snapshot. This makes the interpretation immediate. At \(z=0\), the BCG and central diffuse component are spatially distinct by construction; by \(z\sim0.3\) and especially by \(z\sim1\), however, the stars that will eventually populate the \(z=0\) BCG, ICL, satellite galaxies, and group-scale diffuse component are spread across multiple distinct progenitor systems. In other words, the final central stellar structure is assembled from material that was previously partitioned among several galaxies and groups, rather than simply evolving in-situ inside one long-lived central body.

It is also clear from Figure~\ref{fig:hindcast_split_z0} that the BCG population does not form out of many small galaxies. The orange points are confined to a small number of large progenitors at earlier times. Further, the ICL and group ICL components are diminished but still present at earlier times, with some indication of admixture between the galaxy and ICL populations. 

To quantify which infalling systems matter most, we next identify all mergers onto the main branch. At each snapshot, we find distinct haloes whose descendants either become the next main-branch object directly or become substructure within it one snapshot later. For each such event, we define the merger mass ratio,
\begin{equation}
\mu \equiv \frac{M_{\rm host,max}}{M_{\rm sat,max}},
\label{eq:massratio}
\end{equation}
using the \emph{maximum} mass attained along the previous history of both the host and the satellite, rather than the instantaneous masses immediately before coalescence. This is now standard practice in merger studies, since satellites can lose substantial mass prior to final coalescence and instantaneous masses therefore tend to underestimate the true importance of the merger. For galaxies, \citet{RodriguezGomez2015} showed that the most appropriate merger mass ratio is obtained by evaluating the progenitor masses when the secondary reaches its maximum stellar mass. On the halo side, infalling systems typically reach their peak mass well outside the $R_{200}$ of their eventual host and then begin to lose mass before final merging \citep{Behroozi2014}. Using peak or maximum historical mass therefore provides a more stable proxy for the physical significance of the merger than using the stripped instantaneous mass immediately before coalescence. We note that this choice allows the mass ratio to occasionally fall below 1, i.e. the satellite appears larger than the host. 

For each merger event we then compute two quantities from the \(z=0\) particle tags. First, we measure the total \(z=0\) stellar mass contributed by the infalling halo to the cluster, regardless of where those stars end up at the final time. Second, we measure the subset of that mass that ends up within the final BCG aperture. In Figure~\ref{fig:cluster_buildup}, each point corresponds to one infall event into the cluster (we show cluster 2, the most massive cluster with aligned tracking, see Section~\ref{sec:tracking_clean} above), plotted as a function of lookback time and merger mass ratio. The symbol area scales with the entire stellar mass contributed by that event regardless of where this ends up, while the colour indicates the fraction of the final \(z=0\) BCG aperture mass contributed by the same event. The vast majority of mergers contribute zero mass to the BCG aperture, with only a handful together contributing more than half of the central stars.

Also shown in Figure~\ref{fig:cluster_buildup} is a reference dynamical-friction timescale based on the fitting formula of \citeauthor{BoylanKolchin2008} (\citeyear{BoylanKolchin2008}; Eq.~\ref{eq:bk08} above). We adopt the standard parameter values given above, and show both a central track corresponding to \(\eta=0.5\) and a broader band spanning typical orbital circularities \citep{Wetzel2011}. Mergers that lie well above the curve are rapid enough that a substantial fraction of their stars will be deposited into the central region, while those well below the curve are increasingly unlikely to do so efficiently. Some of the merger events shown in Figure~\ref{fig:cluster_buildup} are marked with green open circles. These events contribute to the BCG component in our forecasting approach introduced in Section~\ref{sec:forecast}.

Although there are a large number of accretion events, the pie chart in Figure~\ref{fig:bcg_buildup_pie}, where the slices represent the ordered fraction of the final BCG aperture stellar mass brought in by each merger, shows that the central region is typically assembled from a handful of important progenitors. The vast majority of minor events contribute little or nothing. Figure~\ref{fig:bcg_buildup_pie} includes a segment labelled `in-situ', indicating the mass fraction of stars that were born belonging to the cluster. In \textsc{GIZMO} these stars always form close to the cluster centre \citep{Contreras2025}. The tracking of the in-situ population is started at a lookback time of 12\,Gyr, $z=3.5$. All the stars that belong to the cluster at this time are included.  Some of the infalling galaxies and groups retain gas that can form additional stars. Stars formed within infalling objects are tracked and accounted for, being added to the infalling stellar mass of the corresponding infalling halo. Via these various channels all of the stars present within the central region at $z=0$ are accounted for. We have further checked that both tracking the stars in the central region at $z = 0$ outwards and following stars crossing $R_{200}$ inwards produces the same set of accretion events and that no events are missed.

Due to the relative importance of a small number of infall events this process is inherently stochastic in nature. To evaluate the average build-up we repeat the above analysis, with Figure~\ref{fig:bcg_average} reproducing Figure~\ref{fig:cluster_buildup} but for the 107 clusters in which the same central object is tracked across the history of the cluster back to a look-back time of 10\,Gyr ($z=1.77$, see Section~\ref{sec:tracking_clean} for details).  As expected, the dynamical friction timescale indicates infall events that are likely to eventually lead to a merger with the cluster centre. Conversely, large mass ratios do not lead to mergers and many of these galaxies survive within the cluster until the present day, typically losing around 30 percent of their stellar mass to stripping \citep{Brown2024}.

We rank the infall events separately for each cluster according to the stellar mass that they contribute to the final BCG aperture, and then average the fractional contribution at each rank across the 107 clusters. Each cluster therefore contributes equally to the mean, irrespective of its total halo or stellar mass. The first four ranked events contribute 24.1, 15.1, 10.4, and 7.6 per cent on average, respectively. The first three together account for 49.6 per cent of the final BCG-aperture stellar mass, while the first four account for 57.2 per cent. The characteristic scale of approximately four events therefore emerges directly from the cumulative ranked contributions.

There is substantial object-to-object scatter around this ensemble average. Defining $N_{50}$ as the minimum number of ranked infall events required to contribute 50 per cent of the final BCG-aperture stellar mass, we find a median $N_{50}=3$, with a 16--84 percentile range of 3--5. Approximately 50 per cent of clusters have $N_{50}\leq3$, 70 per cent have $N_{50}\leq4$, and 84 per cent have $N_{50}\leq5$; the full range is 2--13 events. Thus, while individual assembly histories vary substantially, the majority of BCGs are built from only a handful of dominant infall events.

We find no significant evidence that this result is driven by the most massive clusters. The Spearman correlation between $N_{50}$ and the final host $M_{200c}$ is $r_{\rm s}=0.125$ ($p=0.199$), while the cumulative fraction contributed by the four largest events, $F_4$, has $r_{\rm s}=-0.133$ ($p=0.171$). Dividing the sample into three equal-number halo-mass bins gives median $N_{50}$ values of 3, 4, and 4 from low to high mass. Any mass dependence is therefore weak compared with the object-to-object variation. The ranking itself introduces no additional threshold: events are ordered solely by the stellar mass that they actually contribute to the final BCG aperture. Ranking by another quantity, such as progenitor halo mass, would therefore address a different question from the one considered here.

Figure~\ref{fig:bcg_average_pie} summarises the mean ranked stellar-mass contributions of these infall events. On average around $10\%$ of the mass of the BCG aperture is built up from stars formed in-situ within the centre of the cluster. Around $15\%$ of the mass within the BCG aperture is contributed by numerous small events but the vast majority of infall events contribute nothing to the central region because the dynamical friction timescale is significantly longer than the age of the Universe. Figure~\ref{fig:cluster_buildup} made it clear that much of the BCG aperture stars come from infall events that have mass ratios under 10, i.e. major mergers, that took place four or more Gyr ago, allowing sufficient time for dynamical friction to occur.
As stated, both Figures~\ref{fig:bcg_average} \& \ref{fig:bcg_average_pie} stack 107 of the 324 clusters. While the omitted clusters are generally indistinguishable from those retained, including them impacts both figures: Figure~\ref{fig:bcg_average} has artificial edges where the tracking breaks, as visible to the top right of the figure above but at shorter lookback times. The full pie chart exhibits a shift of $\sim 5\%$ from the in-situ slice to the first halo slice, as broken histories remove material that forms in place and introduces an apparent major merger at the point of the break. Otherwise, the conclusion that the BCG aperture region is built from a small number of major merger events remains.

The same halo infall tracking procedure was also applied to a lower-mass target halo, chosen to be close to \(10^{14}\,{\rm M_\odot}\), in order to test whether the inferred assembly pattern is peculiar to the most massive objects or extends to group-scale masses. The resulting merger mass ratio distributions and pie charts are qualitatively self-similar, indicating that the hindcast picture is not unique to large galaxy clusters but persists into lower-mass central systems at least down to the group scale. We have not studied this recursive behaviour further here. We also note that the same qualitative pattern also appears in other cluster simulation datasets we \citep{HBrown2026} have access to; Horizon-AGN \citep{Dubois_Horizon}, TNG100 \citep{Nelson2019} \& Hydrangea \citep{BaheHydrangea}. 

\subsection{Physical expectation for the BCG--ICL split}
\label{subsec:physicalsplit}

These results suggest a simple qualitative picture; the BCG region is built mainly from the superposition of stars formed in-situ and the dense central stellar systems of a small number of massive progenitors for which dynamical friction is efficient and the available time is long enough for merger completion. The ICL, in contrast, is built from stars that are stripped from lower-mass systems and from the outer regions of more massive progenitors during infall.

This expectation is aligned with \citet{Martin2026}, who find that the phase-space properties of stripped stars and dark matter converge for satellite-to-host mass ratios more disparate than approximately 1:10--1:20. Beyond this point, orbital decay becomes too weak to separate the two components: dynamical friction no longer efficiently delivers stellar cores to the cluster centre, and stripped stars instead trace the cluster halo.

This immediately explains why the BCG and ICL can be physically distinct while remaining spatially interleaved. They are not separated by a sharp edge in radius. Rather, they correspond to different assembly channels: one associated with the efficient incorporation of infalling central galaxies into the bottom of the potential well, and the other with partial stripping, incomplete assimilation, and the broader dispersal of stellar debris through the cluster halo.  Section~\ref{sec:hindcast} showed that only a small number of mergers build up the majority of the stars in the central region of the cluster. This suggests that the assembly history of the dominant progenitors contains sufficient information to identify the material most likely to become part of the final BCG. Section~\ref{sec:forecast} turns this empirical result into a predictive model, using the properties and subsequent dynamical fate of infalling progenitors to forecast the division between the BCG and ICL.

\begin{figure}
\includegraphics[width=\columnwidth]{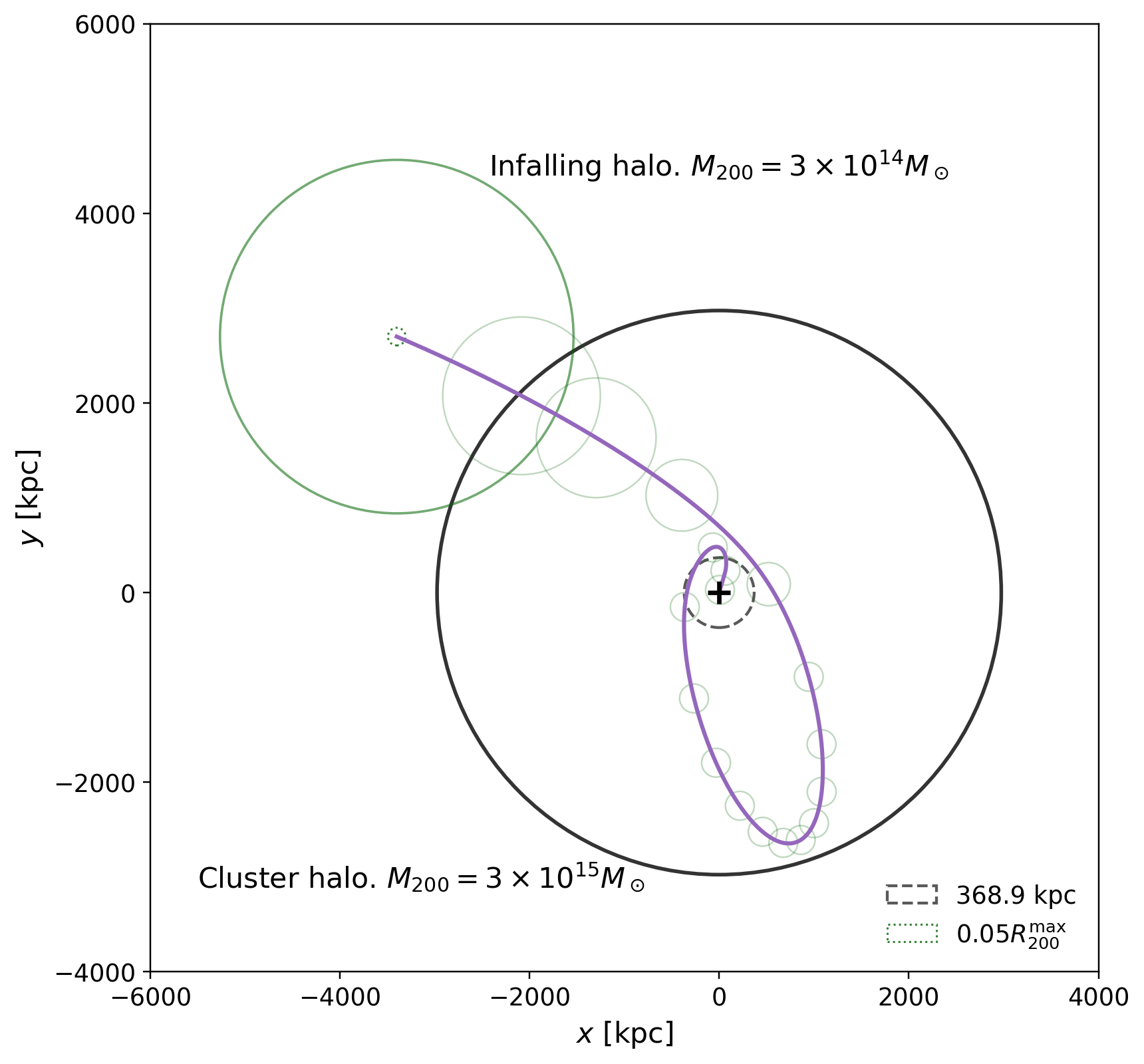}
\caption{Illustration of the forecasting model in action. An infalling halo with a merger mass ratio of 10 suffers both dynamical friction and tidal stripping as it orbits within the cluster. If the halo penetrates within the inner dashed radius and subsequently disappears into the main halo, stars lying within the BCG progenitor central aperture $0.05R_{200}^{\rm max}$ at infall, together with stars forming within this aperture after infall, are assigned to the BCG component. Here $R_{200}^{\rm max}$ is the maximum $R_{200}$ attained by the infalling progenitor along its resolved history. Lighter green circles indicate the stripped radius of the infalling halo at equal time intervals along the orbit.}
\label{fig:orbit_illustration}
\end{figure}

\begin{figure}
\includegraphics[width=\columnwidth]{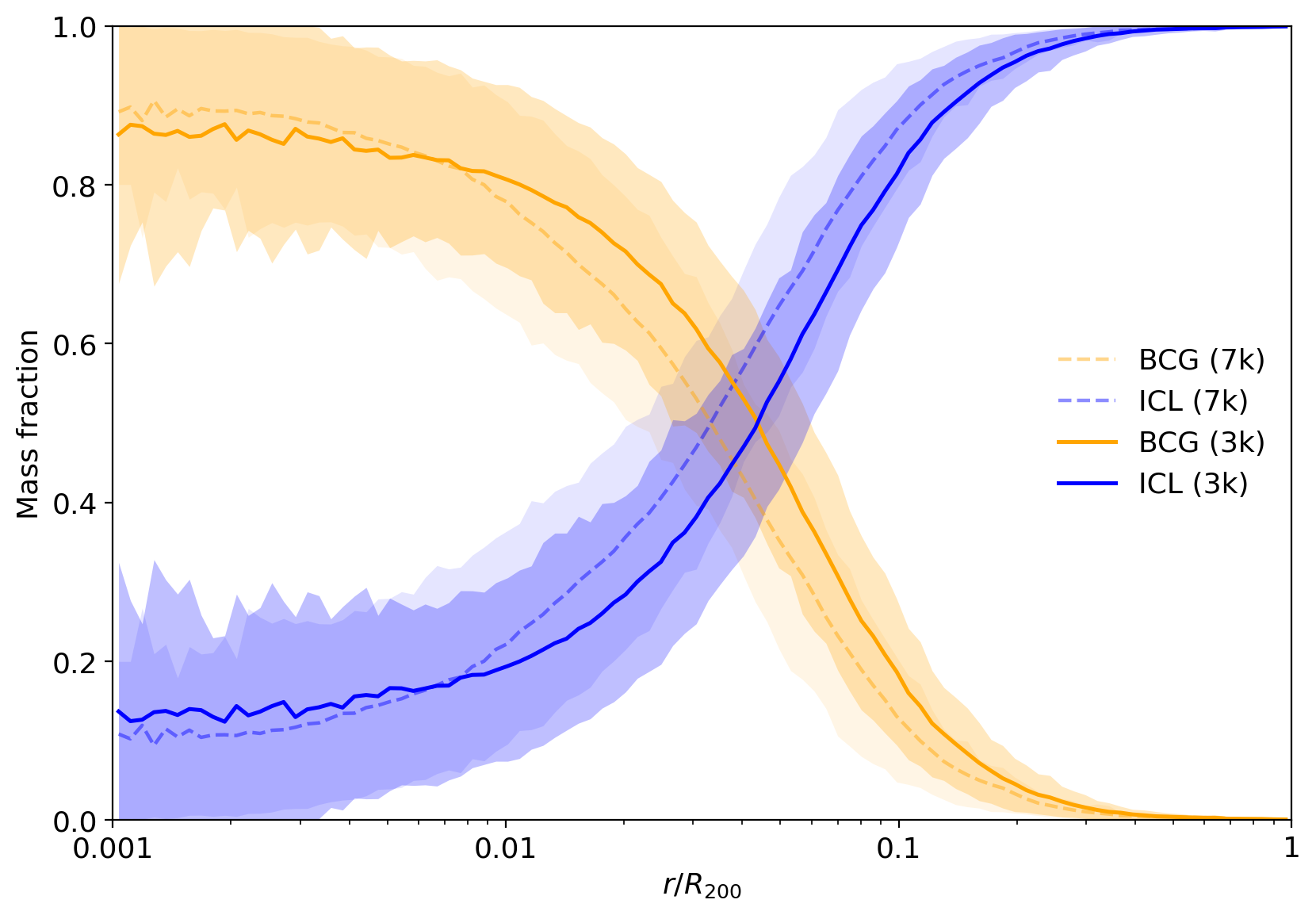}
\caption{Stacked radial mass fractions of the cluster stars, split into BCG and ICL populations, as a function of radius in units of $R_{200}$. Radii are computed relative to the mass-weighted centre of the in-situ stellar component in each cluster. For each cluster, the mass in each category is measured in logarithmic radial bins and divided by the total mass of the cluster stars in the same bin. Solid and dashed curves show the mean stacked fraction across the cluster sample, while the shaded bands indicate the 15th–85th percentile range. Results for the 45 available higher resolution \textsc{GIZMO\_7k} runs are shown as dashed curves with fainter shading.}
\label{fig:forecast_fractions}
\end{figure}

\begin{figure}
\includegraphics[width=\columnwidth]{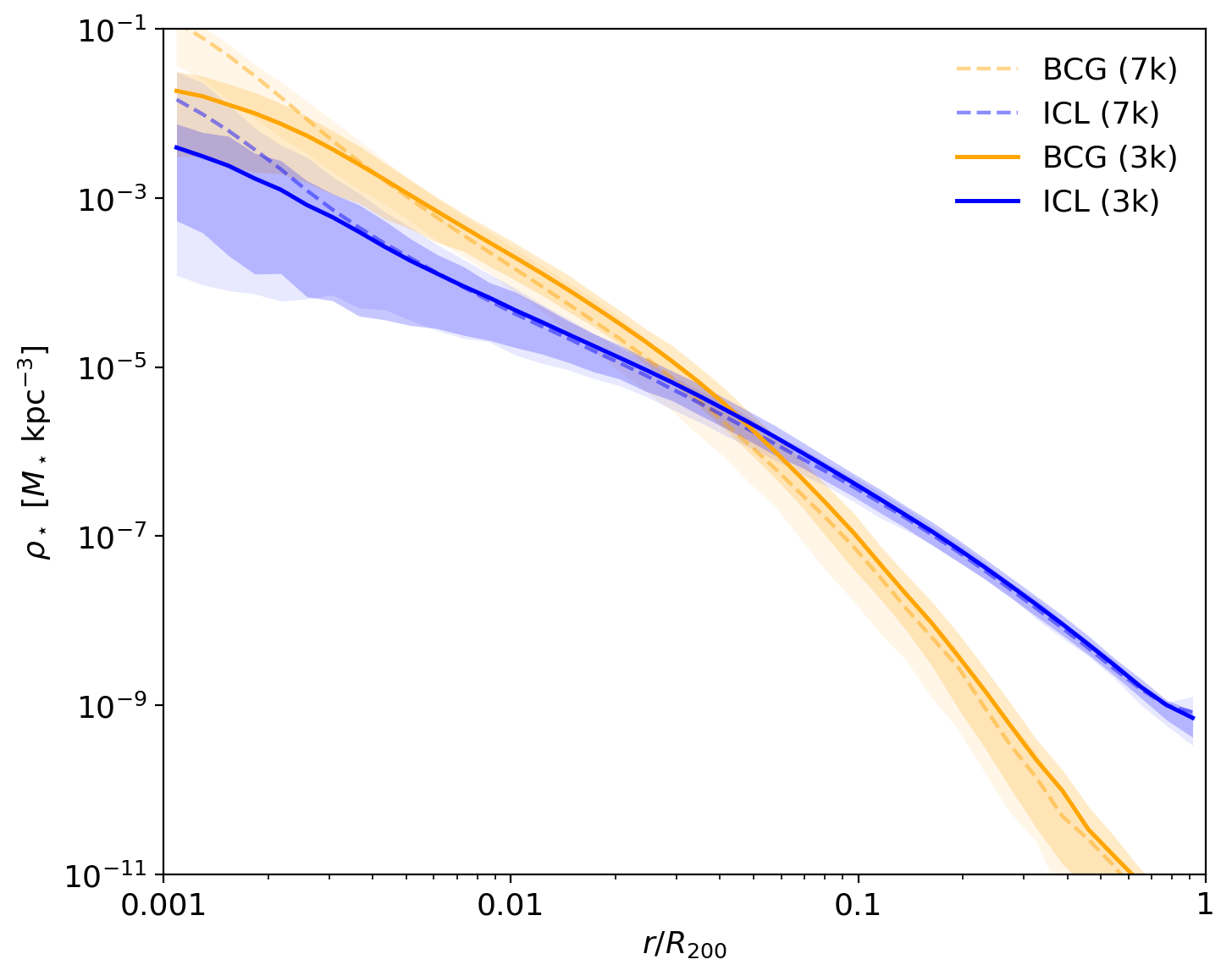}
\caption{Stacked radial density profiles of the BCG and ICL populations, as a function of radius in units of $R_{200}$. Solid and dashed curves show the mean stacked fraction across the cluster sample, while the shaded bands indicate the 15th–85th percentile range. Results for the 45 available higher resolution \textsc{GIZMO\_7k} runs are shown as dashed curves with fainter shading.}
\label{fig:radialdensity}
\end{figure}

\section{A forecasting model for BCG formation}
\label{sec:forecast}

Our aim is to partition the stars contributing to the final central BCG/ICL complex using resolved halo memberships and merger-tree ancestry, rather than by imposing an \emph{a posteriori} radial or kinematic decomposition. We begin
by taking the history of all 324 cluster haloes exactly as constructed above. We drop the four clusters for which the halo history does not extend to $z=3.5$. As previously, we define the set of stars exclusively belonging to each halo, i.e. a star is only a member of the least massive object it is bound to. Even for an isolated halo a small number of stars 
can be lost if they move out beyond $R_{200}$ because this radius does not define a sharp boundary. The fractional mass lost via this channel is always small compared to the mass of stars considered and is anyway not relevant for a discussion of the central BCG/ICL structure.

We first construct an in-situ component, built directly from stars that form within the cluster halo itself rather than within a subhalo of the cluster. We choose to initiate the in-situ component at $z = 3.5$ as prior to that time the halo history breaks up into many small branches and tracking becomes arbitrary. At $z = 3.5$, when the cluster is always small compared to the final object, we label all the stars belonging to the cluster halo at that time as 'in-situ'. We then add to this collection any stars that form within the cluster halo at later times. In \textsc{GIZMO} in-situ stars always form close to the cluster centre. 

We then define a population of infalling systems that make deep passages into the cluster centre and subsequently disappear as independent objects, as illustrated in Figure~\ref{fig:orbit_illustration}. At each snapshot we identify subhaloes of the main cluster that are about to be lost, merging into the main cluster halo. The tracks of these candidate systems are then traced backwards through the merger tree. We retain only those tracks for which the halo contains more than 100 star particles at some stage and for which the minimum separation from the cluster centre falls below $368.9\,\mathrm{kpc}\ ( 250\,h^{-1}\,\mathrm{kpc})$.
For each such progenitor we define $R_{200}^{\rm max}$ as the maximum value of $R_{200}$ attained along its resolved history. This provides a characteristic size for the progenitor before its halo has been substantially reduced by tidal stripping, and avoids defining the stellar selection using an instantaneous radius that shrinks during infall. We then define a BCG progenitor central aperture
\begin{equation}
    r_{\rm cen}=f_{\rm cen}R_{200}^{\rm max}.
\end{equation}
Stars lying within $r_{\rm cen}$ of the progenitor centre when the halo first crosses the cluster $R_{200}$, together with stars that subsequently form within this progenitor-centred aperture, are tagged as belonging to its central stellar component.

If the progenitor subsequently penetrates the central 368.9~kpc region and disappears into the main halo, these tagged central stars are added to the BCG component. The selection therefore identifies stars originating in the compact cores of infalling systems that have undergone sufficient orbital decay to reach the cluster centre before disruption. Stars stripped from the progenitor outside the tagged central component are not subsequently reassigned to the BCG simply because they pass through the central region: once stripped, they orbit independently in the cluster potential and form part of the complementary diffuse component.

This choice has three parameters: the central cluster radius of $368.9\,\mathrm{kpc}$, an imposed minimum star particle count of 100 and $f_{\rm cen}$, which controls the fraction of a BCG progenitor stars that are retained. The central cluster radius is picked to match the maximum distance an infaller can move between output snapshots. As illustrated in Figure~\ref{fig:orbit_illustration} we only have outputs at a finite number of times, roughly $300\,\mathrm{Myr}$ apart (the same separation as the marked circles on the illustration). We tested to check that objects could not move from outside our radius, traverse the core region and exit again between snapshots. A larger choice is not advised because objects can disappear at large radii via mergers within galaxy groups and these are not the events we seek. Later matching between our forecast and hindcast approaches (Figure~\ref{fig:BCGfrac}) reveals one cluster that crosses the central cluster region between two snapshots and then disappears before the next one and is therefore missed. The requirement that a candidate progenitor contains more than 100 star particles at some stage in its resolved history is a numerical resolution criterion rather than a physical mass cut. It ensures that the stellar content of the progenitor is sufficiently well sampled for particle-based tracking and that its merger-tree history can be followed robustly. Lower-mass systems are much more numerous, but individually contribute very little stellar mass to the central BCG \citep{Brown2024, HBrown2026}. In practice, as Figure~\ref{fig:forecast_fractions} reveals, these events can at most contribute only $10\%$ of the central mass given our high completeness inside $1\%$ of $R_{200}$. The final parameter, $f_{\rm cen}$, controls how much of the central stellar component of each qualifying progenitor is retained. It therefore parametrises how deeply embedded the selected stars are within the progenitor before infall. Setting $f_{\rm cen}=0$ recovers the purely in-situ population, whereas $f_{\rm cen}=1$ assigns all stars belonging to a qualifying infalling halo to the BCG component, including its loosely bound outer stellar material.

We explicitly explore the dependence on this parameter over the full range $0\leq f_{\rm cen}\leq1$ in Appendix~\ref{App:parameter} and Figures~\ref{fig:partest_fractions} \& \ref{fig:partest_density}. For the example cluster, reducing the aperture below $f_{\rm cen}=0.05$ causes the fraction of stars in the central region assigned to the complementary ICL component to rise to approximately 10 per cent. We therefore adopt $f_{\rm cen}=0.05$ as our fiducial choice. This value is determined from the BCG progenitor central aperture test and is not chosen to reproduce the final cluster-centric BCG--ICL transition discussed below.

We intentionally do not impose a mass ratio or timescale limit as might be suggested by the theoretical curve of Figure~\ref{fig:cluster_buildup}. While these can be chosen for any particular cluster the process is highly stochastic and including a small fraction of incomplete mergers heavily distorts any stacked result because the radial stellar density drops very rapidly. Our selection ensures that the final set does not include any incomplete mergers. The green circles shown in Figure~\ref{fig:cluster_buildup} indicate the complete list of infall events that qualify for this particular cluster. The low central ICL mass fraction of around $10\%$ seen in Figure~\ref{fig:forecast_fractions} indicates that this approach is highly successful at isolating central stars that either formed in place or came in via objects that suffered significant dynamical friction and underwent disruption.

The remaining stars belonging to the $z=0$ cluster are assigned to the ICL component. These are stars in the $z=0$ cluster halo that have been stripped from infalling objects. The stacked fraction of cluster stars that reside in the BCG and ICL components as a function of radius is shown in Figure~\ref{fig:forecast_fractions}. Of the 324 clusters in \textsc{The Three Hundred} sample we omit four because the tracked history of the main cluster halo does not extend to $12\, {\mathrm {Gyr}}$. We further omit 56 systems in which the identity of the central galaxy swaps at late times, leaving two possible central galaxies. A final three systems are excluded because the $z=0$ centre of mass of the BCG component lies more than half the in-situ aperture (73.8~kpc) from the nominal cluster centre returned by AHF. This leaves 261/324 clusters in the final stack. At $z=0$, these systems span $M_{200c}=7.51\times10^{14}$ to $3.97\times10^{15}\, {\rm M}_{\odot}$. The analysis therefore samples exclusively the high-mass cluster regime, rather than groups or lower-mass systems. The mean values for each population are shown together with the $15^{\rm th}-85^{\rm th}$ percentile range.

Near the centre the BCG component dominates, contributing $80\%$ or more of the stars out to around $1\%$ of $R_{200}$, beyond which the fraction rapidly declines. Outside of $5\%$ of $R_{200}$ the ICL group dominates, originating from stars stripped from surviving groups and galaxies. We underplot as dashed curves and fainter range shading similar results from a stack of available \textsc{GIZMO\_7k} cluster simulations that form part of a later convergence test. The results are very comparable with a trend to smaller radii for the crossover between the two populations.

In Figure~\ref{fig:radialdensity} we convert these radial fractions into the stellar density profiles of the two components, again with results from both \textsc{GIZMO\_3k} (solid lines) and \textsc{GIZMO\_7k} (dashed lines) included. The stack is for the same set of clusters as displayed in Figure~\ref{fig:forecast_fractions}. The BCG component dominates the ICL component within around $4\%$ of $R_{200}$ ($3\%$ for \textsc{GIZMO\_7k}).

It is important to distinguish this cluster-centric transition scale from the BCG progenitor central aperture introduced above. The fiducial selection $r_{\rm cen}=0.05R_{200}^{\rm max}$ is defined separately for each infalling progenitor, using the maximum $R_{200}$ attained by that progenitor. By contrast, the radial coordinate in Figures~\ref{fig:forecast_fractions} and~\ref{fig:radialdensity} is normalised by the $R_{200}$ of the final $z=0$ host cluster. The appearance of a BCG--ICL crossover at a few per cent of the final host $R_{200}$ is therefore not imposed by the choice $f_{\rm cen}=0.05$.

The numerical similarity is nevertheless intriguing. In a different context, \citet{ContiniYi2026} identified an optimal host-centred BCG--ICL separation aperture of $r_{\rm cut}/R_{\rm vir}\simeq0.045$ around a halo mass of $10^{14}\,{\rm M}_{\odot}$. Their radius is defined relative to the host halo and therefore differs fundamentally from our BCG progenitor central aperture. We consequently do not interpret the agreement as evidence that these quantities represent the same physical scale. It may, however, indicate that scales of a few per cent of a halo virial radius commonly mark the transition between a compact stellar component associated with the central potential and a more extended population increasingly governed by the halo potential. Establishing whether such a scaled transition is universal would require a direct comparison using consistent definitions.

\begin{figure}
\includegraphics[width=\columnwidth]{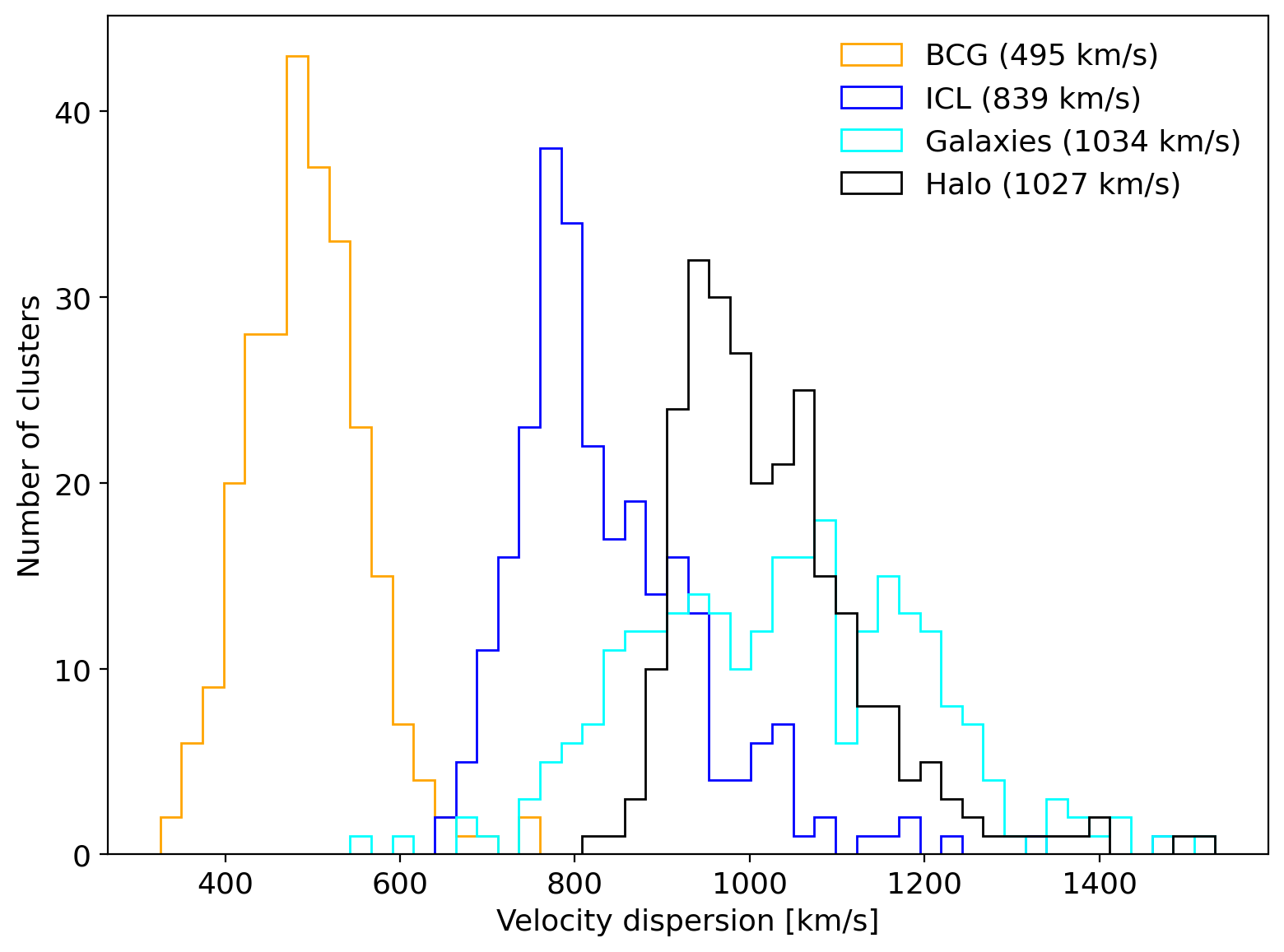}
\caption{Distribution of one-dimensional velocity dispersions for the in-situ, BCG, ICL, galaxy, and main-halo dark matter components across the cluster sample at $z=0$. The stellar dispersions are computed from the mass-weighted internal kinematics of the corresponding particle sets, while the galaxy dispersion is derived from subhalo bulk velocities. The mean velocity dispersion of each population is given in the legend.}
\label{fig:dispersion_histogram}
\end{figure}

\begin{figure}
\includegraphics[width=\columnwidth]{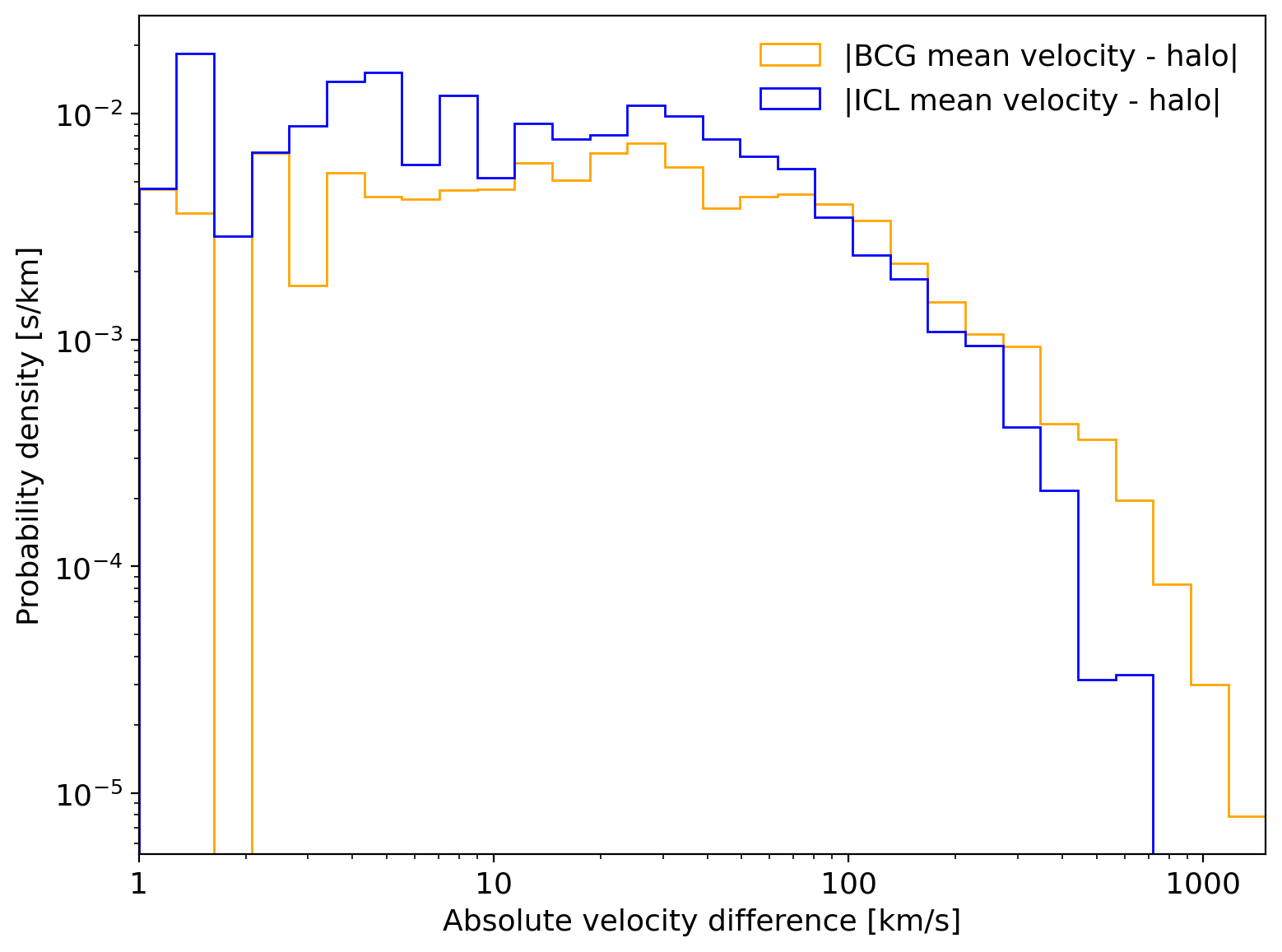}
\caption{Distribution of mean stellar velocity offset relative to the main halo for the BCG and ICL populations at 
$z=0$. For each cluster, the mass-weighted mean velocity of each stellar component is computed and the halo bulk velocity subtracted; the histogram then shows the distribution of the resulting velocity differences.}
\label{fig:vdiff}
\end{figure}

\begin{figure*}
\centering
\includegraphics[width=0.49\textwidth]
{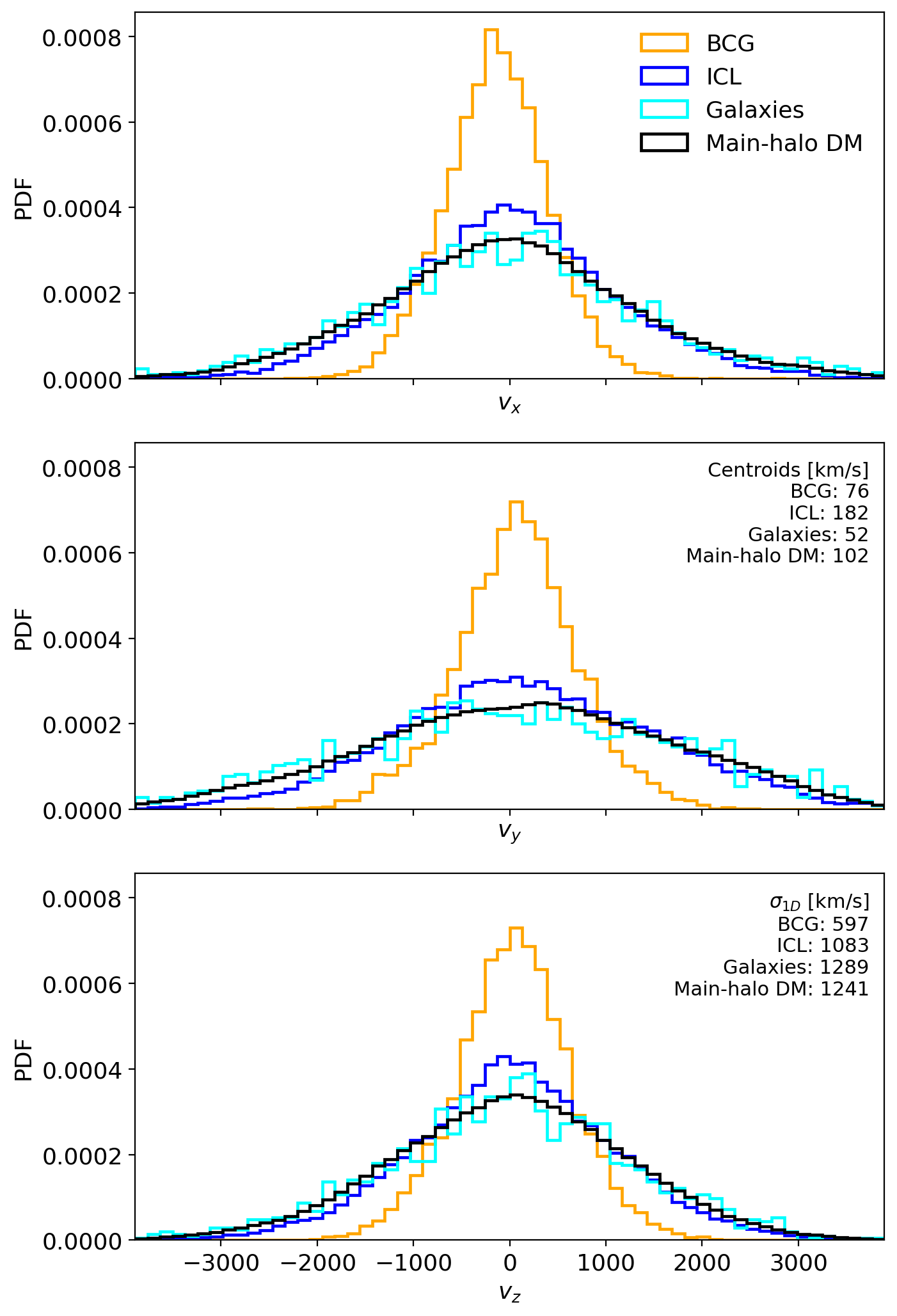}
\hfill
\includegraphics[width=0.49\textwidth]{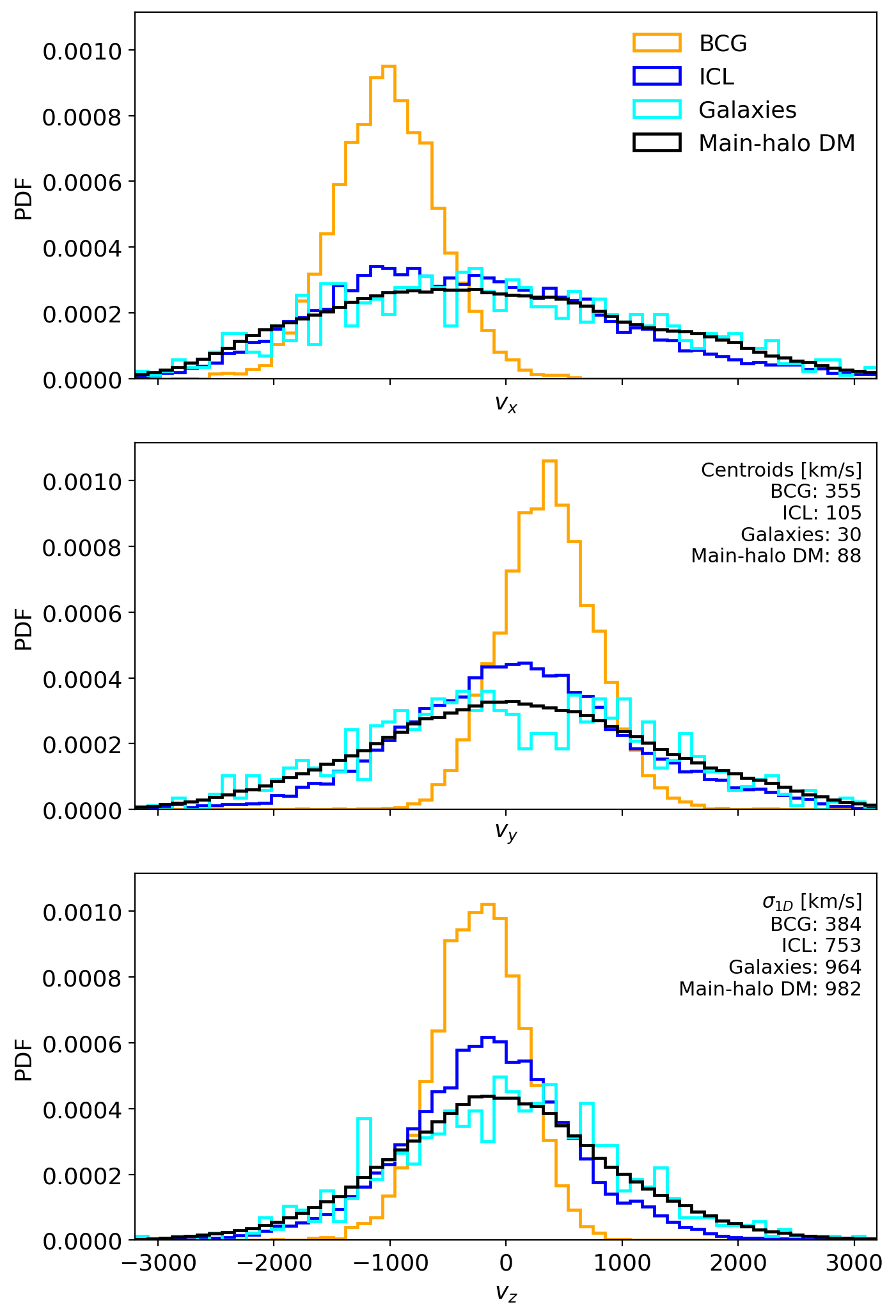}
\caption{Probability density functions of the cluster rest-frame velocity components of the main stellar and halo populations for cluster 2 at $z=0$ (left column). Shown are the BCG and ICL stellar components, together with the velocities of the galaxies and the main-halo dark matter. All histograms are normalised to unit area. The corresponding velocity centroids and velocity dispersions for each distribution are given in the lower two panels, respectively. The right column is for cluster 57, where the central BCG complex is offset in velocity space from the velocity distribution of the full halo. }
\label{fig:cluster_velocity_hist}
\end{figure*}

\subsection{Velocity-space separation of the central stellar components}
\label{subsec:velocity_space}

\begin{figure}
\includegraphics[width=0.49\textwidth]{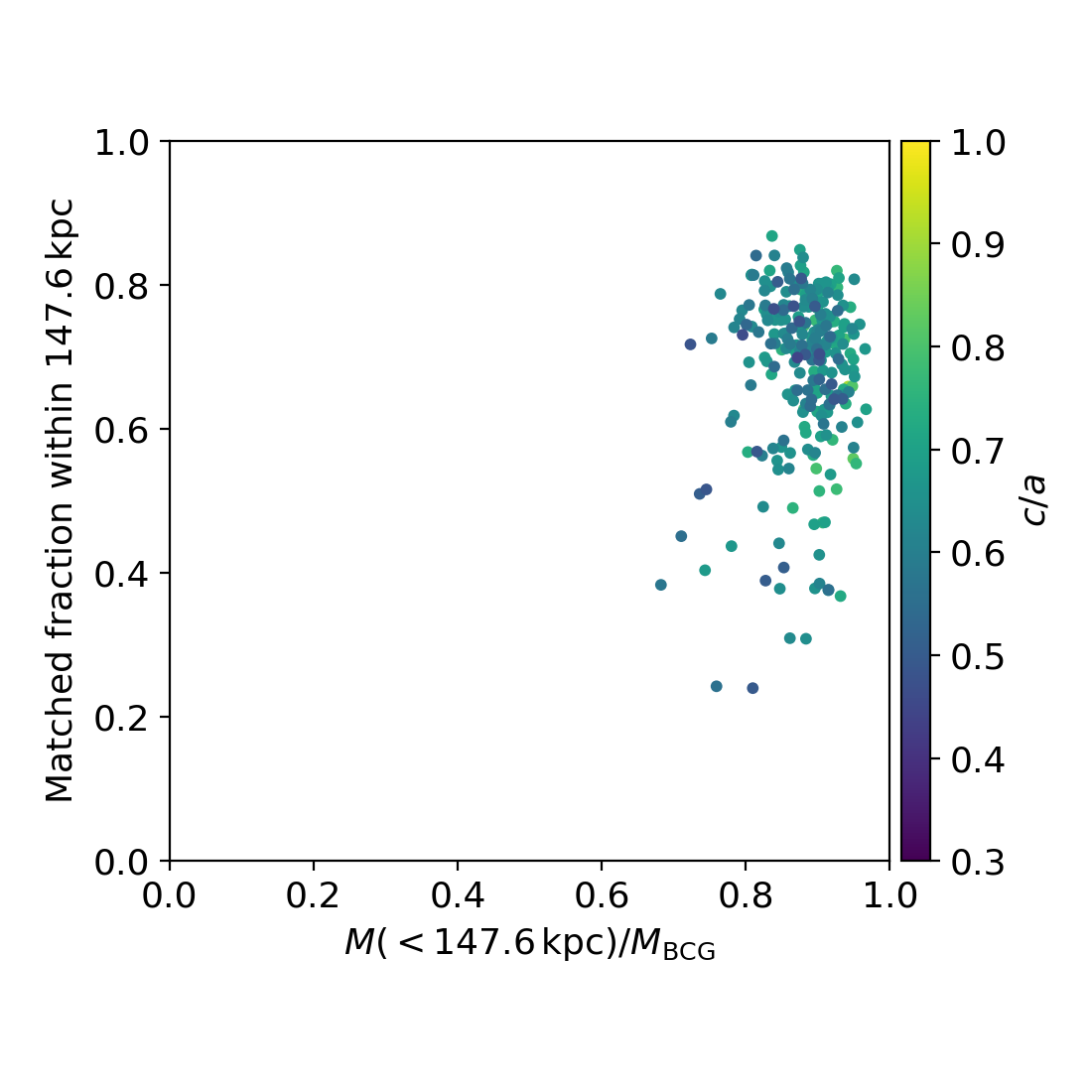}
\caption{Comparison of the aperture completeness and concentration of the physically defined central stellar component. The vertical axis shows the fraction of the cluster stellar mass within $147.6\,\mathrm{kpc}$ that belongs to the combined BCG component. The horizontal axis shows the fraction of the mass of the BCG component that lies within the same aperture. Points are colour-coded by the minor-to-major axis ratio, $c/a$, of the BCG stars within $147.6\,\mathrm{kpc}$.}
\label{fig:BCGfrac}
\end{figure}

A complementary view of the stellar spatial decomposition is provided by the velocity distributions of the different stellar and halo populations at $z=0$. For each cluster we measure the one-dimensional velocity dispersions of the BCG and ICL stellar components of the final cluster, together with those of satellite galaxies and the dark matter of the cluster itself. Figure~\ref{fig:dispersion_histogram} shows the distribution of these dispersions across the 261 clusters we retained from the \textsc{GIZMO\_3k} sample, with the mean velocity of each group given in the legend. The BCG component is the dynamically coldest, with an average velocity of around  $500\,\mathrm{km\,s^{-1}}$.The ICL population is much hotter, sitting at around $840\,\mathrm{km\,s^{-1}}$. The galaxies and the parent halo have similar temperatures around $1000\,\mathrm{km\,s^{-1}}$ but the mass weighted galaxy distribution is significantly broader. This ordering is exactly what would be expected if the BCG component traces the deepest central potential. The BCG component represents stars deposited by massive infallers that have sunk well into the cluster core, losing energy via dynamical friction, while the ICL corresponds to a more extended and dynamically hotter diffuse stellar population stripped from orbiting galaxies having lost much less energy prior to being stripped.

Figure~\ref{fig:vdiff} provides a different diagnostic by showing the distribution of the mean velocity offsets of the BCG and ICL components relative to the mass-weighted bulk velocity of all the particles belonging to the main halo. Both populations are generally centred close to the halo rest frame, but with a non-negligible scatter from cluster to cluster of around $100\,\mathrm{km\,s^{-1}}$. 
The potential for an offset between the BCG component and the cluster halo is important: the central stellar population is not always perfectly locked to the bulk halo velocity, and the degree of offset carries information about the recent assembly state of the cluster \citep{Behroozi2013}.

To illustrate the underlying distribution functions directly, Figure~\ref{fig:cluster_velocity_hist} shows the cluster rest frame velocity probability density functions for two example systems. In each case we plot the three Cartesian components after subtracting the bulk velocity of the cluster. Shown are the BCG and ICL stellar components, together with the bulk velocities of the galaxies and the particle velocities of the cluster dark matter. All distributions are normalised to unit area.

The left column of Figure~\ref{fig:cluster_velocity_hist} shows the behaviour in cluster 2, a comparatively regular system. The BCG population is narrow, being significantly colder than the ICL population, whose dispersion remains somewhat lower than that of the surrounding cluster. The galaxies show a broader distribution close to that of the cluster, as expected for orbiting satellites rather than stripped stars. The cluster dark matter broadly outlines the global potential as the majority of the mass is in this component.

The right column of Figure~\ref{fig:cluster_velocity_hist} illustrates a less regular case in which the central stellar complex is visibly offset from the bulk cluster rest frame. Here the BCG distribution remains narrower than the galaxy and dark-matter components, but the centroid is shifted, showing that the central stellar structure sits within its own section of the wider cluster halo.

Taken together, these figures support a broader interpretation of the forecast-defined central stellar decomposition. The BCG component is dynamically cold, as expected for stars deposited by deep, massive infall events. It is substantially colder than the ICL population. The galaxies and dark matter provide useful external references: the former represent the kinematics of surviving orbiting substructure, while the latter essentially forms the overall cluster potential. The fact that the stellar components occupy systematically different regions of velocity space strengthens the conclusion that the decomposition identifies physically distinct assembly channels, even though the corresponding stars remain spatially interleaved.

This point is important observationally. One should not expect a perfectly sharp spatial boundary between the central BCG and the surrounding diffuse stellar component, because stars from different assembly channels overlap strongly in radius. However, the velocity-space separation shows that the overlap is not arbitrary. The transition from the colder central population to the hotter diffuse population occurs in a manner consistent with the different depths to which stars from different progenitors have sunk within the cluster potential. Any observational separation between BCG and ICL will therefore remain approximate, but it is nonetheless worth attempting to capture a real underlying dynamical distinction rather than impose a purely spatial division.

\subsection{Convergence of hindcast and forecast classifications}

A key test of the physical decomposition developed in this work is the degree to which it reproduces the central stellar component identified by the conventional hindcast aperture at $z=0$. We therefore compare the stars classified as belonging to the BCG component in the forecasting framework with the set of cluster stars lying within the fixed $100\,h^{-1}\,{\rm kpc} (147.6\,{\rm kpc})$ aperture introduced in Section~\ref{sec:hindcast}. This aperture is used here only as an external reference and plays no role in the forecast classification itself.

Figure~\ref{fig:BCGfrac} compares the aperture completeness of the physically defined BCG component with its degree of central concentration, colour coded by the minor-to-major axis ratio, \(c/a\), measured from the 3-dimensional inertia tensor. The vertical axis shows the fraction of cluster stellar mass within $147.6\,\mathrm{kpc}$ that is assigned to the BCG component. The horizontal axis shows the converse quantity: the fraction of the total BCG component stellar mass that lies within the same aperture.

Most clusters lie close to the upper-right region of the diagram, indicating that the forecast-defined BCG component is both largely contained within $147.6\,\mathrm{kpc}$ and accounts for most of the cluster stellar mass inside that aperture. The small number of systems with lower completeness are preferentially more extended and have lower values of $c/a$, indicating more elongated central stellar distributions. These objects are therefore not simple failures of the physical classification, but dynamically disturbed systems in which a spherical aperture is a poor description of the central stellar structure. 

The outliers on Figure~\ref{fig:BCGfrac} with low values of matched fraction within the aperture have two sources: 
the majority (8 of the bottom 10) are due to a tracking issue. In all these cases an infalling halo with an appropriate merger mass ratio is missed because the centre of the cluster halo is attached to it on first infall. As this infalling halo is not seen within the central radius of $368.9\,\mathrm{kpc}$ before it becomes the cluster halo it is missed by our algorithm. The remaining two halos are complicated mergers where several infallers fall just below the threshold to be counted as major mergers. Removing these clusters from the stacked sample makes no appreciable difference to any of our results.

We have also tested whether the performance of the forecast classification depends systematically on host halo mass. Both matching fractions shown in Figure~12 exhibit weak negative correlations with $M_{200c}$. For the fraction of stellar mass within 147.6~kpc assigned to the forecast BCG component, we find $r_{\rm s}=-0.127$ ($p=0.041$). For the fraction of the total forecast BCG component contained within the same aperture, we find $r_{\rm s}=-0.216$ ($p=4.3\times10^{-4}$). When the sample is divided into three equal-number mass bins, the corresponding median fractions decrease only from 0.721 to 0.699 and from 0.899 to 0.881, respectively, between the lowest- and highest-mass thirds. These changes are small compared with the cluster-to-cluster scatter, indicating only a weak deterioration in the aperture correspondence towards the highest-mass systems.

Taken together, these results demonstrate that the hindcast and forecast approaches are closely aligned for relaxed systems, and that deviations arise primarily in dynamically complex environments where no simple geometric definition of the BCG is expected to perform well. This provides strong support for the physical interpretation of the BCG component as tracing the stars that are retained through a handful of major mergers that suffer significant dynamical friction lowering their orbital energy.

\begin{figure}
\includegraphics[width=\columnwidth]{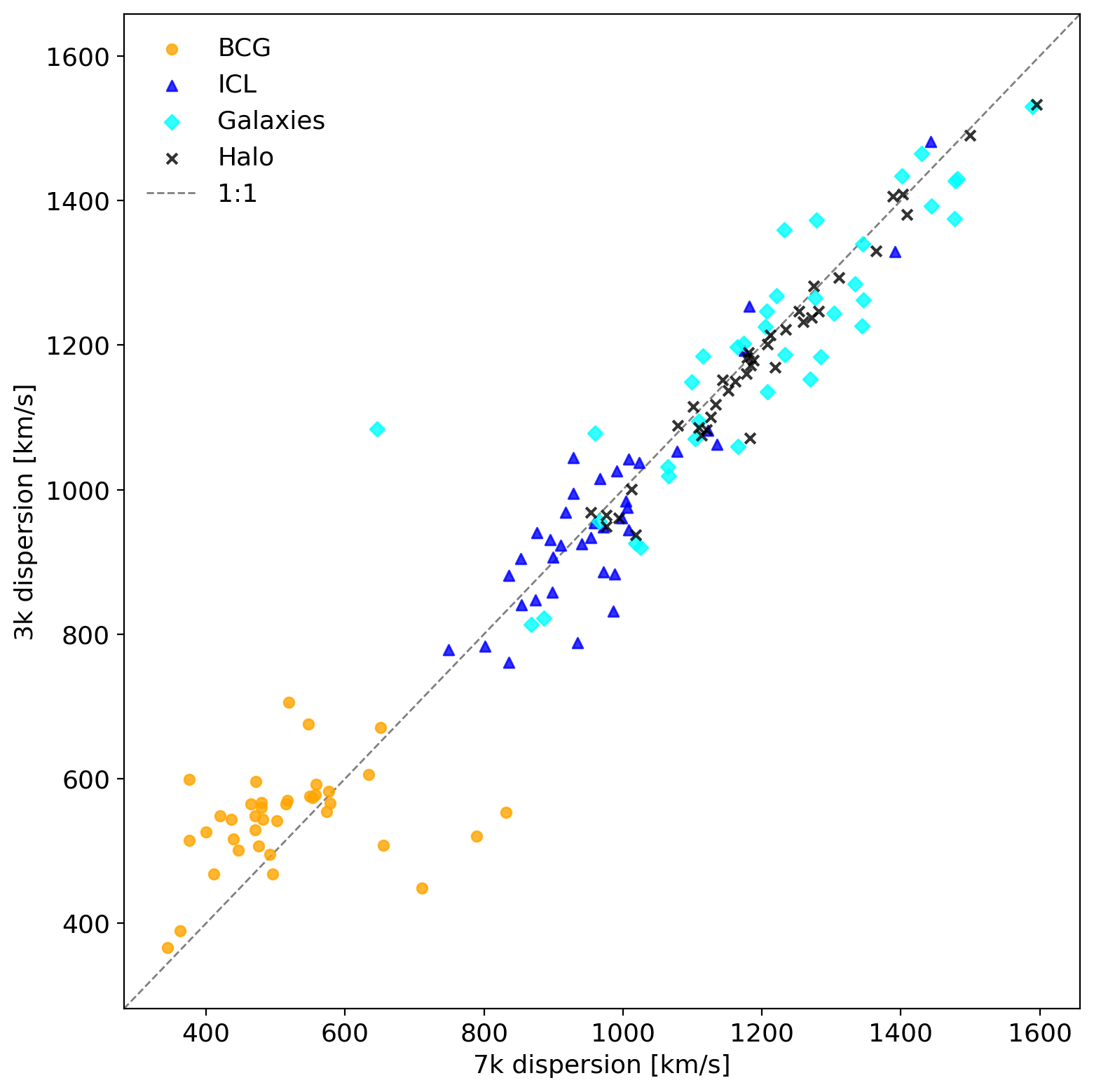}
\caption{Comparison of one-dimensional velocity dispersions at $z=0$ for matched clusters in the \textsc{GIZMO\_7k} and \textsc{GIZMO\_3k} simulations. Shown are the dispersions of the in-situ, BCG, and ICL stellar components, together with those of the galaxies and the dark matter halo. Each point represents a single cluster with the dashed line marking equality.}
\label{fig:dispersion_3k_37}
\end{figure}

\subsection{Impact of resolution}
\label{subsec:resolution}

It is important to check whether the physically motivated decomposition adopted here is robust to finite numerical resolution, particularly because the forecasting model includes a minimum progenitor particle count. In Figures~\ref{fig:forecast_fractions} and~\ref{fig:radialdensity} we therefore compare the results with the 45 available higher-resolution GIZMO\_7k runs.

The same minimum threshold of 100 star particles is applied at both resolutions. Since the GIZMO\_7k particle masses are eight times smaller, this criterion corresponds to an approximately eight-times lower stellar-mass scale and consequently allows substantially smaller progenitors to be followed. The comparison therefore provides a direct test of whether unresolved low-mass accretion events strongly bias the inferred BCG--ICL decomposition. In Figures~\ref{fig:forecast_fractions} and~\ref{fig:radialdensity} we underplotted the results from the 45 available \textsc{GIZMO\_7k} runs as dashed lines with fainter shading indicating the range. The \textsc{GIZMO\_7k} clusters produce a crossover between BCG and ICL dominated regimes that is at a smaller radius but otherwise the profiles are very comparable.  

In Figure~\ref{fig:dispersion_3k_37} we compare the kinematic properties of the different populations taken from matching \textsc{GIZMO\_3k} and \textsc{GIZMO\_7k} clusters. The same scripts and thresholds were used at both resolutions. The particle count thresholds therefore extend to 8 times lower masses at higher resolution and correspondingly more objects are tracked. The ICL, galaxy and dark-matter halo dispersions lie close to the one-to-one relation, indicating that these broader and dynamically hotter components are only weakly affected by the change in resolution. By contrast, the BCG population shows a small systematic offset, with the \textsc{GIZMO\_7k} runs yielding slightly lower dispersions by roughly $50\,\mathrm{km\,s^{-1}}$ on average. In other words, the central stellar populations become somewhat dynamically colder at higher resolution, while the kinematics of the more extended cluster components remain largely unchanged. This is consistent with the crossover of the BCG and ICL radial profiles moving inwards as observed in Figure~\ref{fig:forecast_fractions}.

This behaviour is physically plausible. The BCG stars are the most centrally concentrated stellar population and therefore the most sensitive to the particular star formation and energy feedback model employed as well as the detailed structure of the inner potential. Increasing the resolution allows the gas out of which these stars form to settle into a colder central configuration and form at an earlier time. These stars are precisely the ones most impacted by the resolution change.  The broader ICL and the satellite-galaxy population, by contrast, are controlled more by the global cluster potential and by the orbital distribution of accreted systems, and so are much less sensitive to this change.

The radial comparison in Figure~\ref{fig:forecast_fractions} shows that this kinematic change has only a modest spatial counterpart. The overall stacked decomposition of the \textsc{GIZMO\_7k} runs is very similar to that seen in the \textsc{GIZMO\_3k} runs. The main effect is that the BCG contribution extends to somewhat smaller radii in the \textsc{GIZMO\_7k} runs, as would be expected for a population whose velocity distribution is systematically colder. Beyond this moderate contraction of the BCG component, there is little evidence for a strong resolution dependence in the radial structure of the decomposition.

The similarity of the two resolutions indicates that the unresolved low-mass progenitor population does not dominate either the central BCG assembly or the diffuse ICL structure. Finite resolution can nevertheless affect the detailed balance between these channels: smaller satellites are resolved for longer at higher resolution, and their stellar stripping histories can therefore be followed more completely. Differences in the treatment and numerical implementation of baryonic processes, including cooling, star formation, and feedback, may also influence the detailed stellar distributions and hence the inferred balance between BCG and ICL growth. The principal measurable effect in the present comparison is instead confined to the compact BCG component, whose velocity dispersion is lower by approximately $50\,{\rm km\,s^{-1}}$ and whose radial extent is slightly reduced in the GIZMO\_7k runs.

These results suggest that the physically-motivated classification developed here is reasonably robust to resolution. The main qualitative picture is unchanged: the central stellar component remains decomposable into populations with distinct assembly histories and distinct dynamical states. The principal quantitative sensitivity is confined to the central BCG-like population, which becomes somewhat colder and slightly less extended at higher resolution, while the diffuse ICL and the surrounding cluster populations are largely unaffected.

\section{Implications of a Physical BCG--ICL Decomposition}
\label{sec:implications}

The forecasting analysis changes the interpretation of the BCG--ICL division. In the conventional hindcast approach, the decomposition at \(z=0\) is imposed using a fixed spatial aperture or similar operational definition. While such choices are convenient and observationally motivated, they are not physically unique. Even in three dimensions the transition between the compact central galaxy and the diffuse stellar envelope is continuous rather than sharp, and in projection the components are further blended along the line of sight. Fixed apertures therefore provide a practical definition, but not a fundamentally meaningful one.

The forecasting approach instead classifies stars according to the dynamical fate of their parent systems during hierarchical assembly. In this picture, the relevant question is not where a star lies at $z = 0$, but whether its progenitor experienced sufficient dynamical friction to deliver its stellar content to the deeply bound central region of the cluster potential prior to disruption. Stars remaining within such systems are naturally identified with the central galaxy, while stars stripped from infalling systems remain on more extended orbits and contribute to the diffuse intracluster component.

It is important to distinguish clearly between the evolution of bound systems and that of individual stars. Dynamical friction acts on satellites, reducing their orbital energy and driving them towards the centre. Tidal stripping transfers stars from these systems onto collisionless orbits within the cluster potential. Once unbound, these stars approximately conserve their orbital energies and subsequently phase-mix, rather than undergoing any further orbital decay. The BCG--ICL division therefore reflects how far a system progresses along this dynamical pathway prior to disruption.

The broad assembly picture that emerges here is consistent with previous theoretical studies, despite substantial differences in how the central and diffuse stellar components are defined. Semi-analytic models have long predicted that BCG stars form relatively early but that the galaxy itself assembles substantially later through mergers \citep{DeLucia2007}. Hydrodynamical cluster simulations similarly find late assembly and strong aperture dependence in the inferred BCG growth \citep{RagoneFigueroa2018}. More recently, the TNG300 analysis of \citet{MontenegroTaborda2023} found that approximately 70 per cent of present-day BCG stellar mass is ex-situ, with stars accreted in major mergers preferentially concentrated towards the centre, while stars tidally stripped from surviving galaxies make a much larger contribution to the outer ICL. Our result that the central stellar component is assembled primarily from the dense cores of a small number of massive progenitors, whereas stripped material preferentially populates the extended cluster potential, is therefore qualitatively consistent with results obtained using substantially different simulations and definitions. The distinction in the present work is that these populations are identified from their assembly histories and dynamical fates rather than separated by their final radius.

The relative importance assigned to mergers and tidal stripping in building the diffuse component is less universal. In hydrodynamical simulations, \citet{Murante2007} found that most diffuse stars became unbound during mergers associated with the BCG and other massive galaxies, with stripping contributing only a minority of the diffuse component. By contrast, the semi-analytic models of \citet{Contini2014,Contini2018} find a substantially larger, and in their standard model dominant, contribution from tidal stripping, primarily from intermediate- and high-mass satellites. These results are not directly contradictory because the boundary between `merger' and `stripping' channels is itself definition dependent: stars lost shortly before final coalescence may be classified as merger debris in one scheme and as tidally stripped material in another \citep[see also the discussion in][]{Contini2018,Contini2021}.

Our decomposition makes a particular physical choice in this regard. The compact stellar core of a progenitor that survives orbital decay to the cluster centre is associated with the assembled BCG, whereas stellar material removed from that system before central assimilation is associated with the ICL. The absolute fractions assigned to the two channels should therefore not be compared directly with studies employing different spatial or event-based definitions. The more generic result is the dynamical sorting itself: centrally retained material is preferentially associated with massive progenitors that undergo substantial orbital decay, while more weakly bound material is stripped onto extended cluster orbits.

This does not imply that the BCG and ICL become spatially distinct in any simple morphological sense. On the contrary, the two components remain strongly interleaved. The central galaxy is embedded within a diffuse stellar envelope built from stripped material. The resulting structure is therefore physically meaningful but has no hard boundary, consistent with the long-standing difficulty of separating BCG and ICL observationally using surface-brightness thresholds, profile fitting, or fixed apertures \citep[e.g.][]{Gonzalez2005,Zibetti2005,Cui2014,DeMaio2020,Brough2024}. It is also clear why practical aperture selection methods work so well: a small central window preferentially selects for stars brought in via major mergers, so the two approaches produce broadly similar outcomes.

The stochastic nature of cluster assembly further complicates any structural interpretation. The inner regions are typically built from a small number of major events, and individual systems often display slope changes or apparent edges in their radial profiles that reflect specific merger histories. While such features can be prominent, they are not generic or globally meaningful, and do not correspond to a universal physical boundary between components.

Attempts to predict the outcome of individual accretion events using simple dynamical-friction timescale criteria are only partially successful. Although such approaches broadly identify systems likely to merge, they fail to capture the full complexity of the process. Some systems that satisfy nominal merger criteria retain significant bound remnants at \(z=0\), while others are fully disrupted. Tightening the selection to exclude incomplete mergers inevitably removes genuine events. Moreover, even a clean identification of mergers does not directly yield the subset of stars that are ultimately deposited in the central region without reconstructing orbits and tracking particle histories. Given the relatively small number of important events, even a handful of misclassified systems can dominate the residual signal at intermediate radii.

In contrast, selecting stars associated with disrupted systems on plunging orbits provides a robust empirical indicator that strong dynamical friction has operated. When combined with stars formed in-situ near the bottom of the potential, this yields a stellar population that is highly complete within the central few per cent of \(R_{200}\). This is a notable result given the dynamical complexity of this region.

Most accreted galaxies survive as galaxies, for a substantial fraction of the age of the Universe, and most infall events do not contribute anything to the final BCG. In physical terms, bringing stars into the cluster is not the same thing as delivering them to the centre. Many stars remain in surviving satellites or are stripped into the broader intracluster component without ever becoming part of the compact central galaxy. This is precisely why a purely merger count based view of BCG assembly is inadequate: the key question is not whether an object fell into the host halo, but whether its stars had time and the right dynamical circumstances to reach the central galaxy itself. This insight is incorporated into the forecasting model we have proposed.

The central stellar complex itself is best understood as a superposition of two closely related contributions. The first is a genuinely in-situ population, formed within the central potential well. We intend to study the specific population further in a subsequent paper (Pearce et al., in prep). The second consists of stars formed in progenitor systems, via the same underlying subgrid star-formation physics, that have subsequently been delivered to the centre through strong dynamical friction. These populations are in some sense entirely interchangeable: given two equal mass progenitor objects, which is chosen as the cluster and which is the infaller is an arbitrary choice. Essentially any of the handful of objects building up the BCG component could be viewed as the 'in-situ' cluster. Practically it makes no difference, as all the objects are superimposed on top of each other, becoming indistinguishable. No sign of the original split into multiple stellar populations remains in the core, with the stars of the BCG component being well mixed in phase space.

Within each large halo the in-situ population provides a useful reference scale. Having formed near the bottom of the potential, these stars do not experience subsequent orbital energy loss via dynamical friction, and their spatial extent is therefore relatively fixed. In the \textsc{GIZMO\_3k} runs this central core extends to less than $\sim0.5\%$ of $R_{200}$ ($\lesssim30\,\mathrm{kpc}$). Its velocity dispersion is somewhat higher than observed BCG values \citep[e.g.][]{Loubser2018ICL}. At higher resolution (\textsc{GIZMO\_7k}), the dispersion decreases by $\approx 50\,\mathrm{km\,s^{-1}}$, moving towards observational constraints, indicating some residual resolution dependence and that the compact core is only marginally resolved at our resolution. When a handful of these cores are stacked together we build our BCG component. The mean velocity dispersion of these is $501\, \mathrm{kms^{-1}}$.

Several aspects of this picture are already accessible observationally, although the correspondence with the dynamical classification developed here is necessarily indirect. Deep imaging can measure the projected radial surface-brightness distribution of the BCG+ICL complex, while spatially resolved colours and stellar population modelling constrain how the underlying stellar population changes with radius. Observed negative colour and metallicity gradients in the diffuse light have generally been interpreted as evidence that a substantial fraction of the ICL originates from stars stripped from the outer regions of intermediate- or high-mass galaxies, rather than being produced predominantly by major mergers with the BCG \citep[e.g.][]{DeMaio2015,DeMaio2018,Montes2018}. Such measurements provide a direct observational test of our prediction that stripped material becomes increasingly important with cluster-centric radius. In particular, \citet{Montes2018} infer an ICL metallicity of approximately $[\mathrm{Fe/H}]\simeq-0.5$ in massive clusters, consistent with stripping from the outskirts of roughly Milky-Way-mass systems.

Our kinematic prediction is potentially even more useful: our forecast BCG and ICL populations overlap strongly in radius but have substantially different velocity dispersions. This behaviour can be observed in nearby systems using integrated-light spectroscopy \citep{Kelson2002, Bender2015, Gu2020} and discrete tracers such as planetary nebulae and globular clusters \citep{Longobardi2015, Hartke2022}. Both reveal rising velocity dispersion profiles consistent with an extended ICL component. Furthermore, in M87, \citet{Longobardi2015} separated co-spatial stellar-halo and intracluster populations kinematically, finding a narrow component associated with the galaxy and a broader ICL component. This is qualitatively similar to the separation seen in Figures~\ref{fig:dispersion_histogram}, \ref{fig:vdiff} \& \ref{fig:cluster_velocity_hist}. Measurements of the radial transition in velocity dispersion may therefore offer a more direct observational proxy for the dynamically defined BCG--ICL separation than a single surface-brightness or aperture threshold.

Individual systems can also provide direct snapshots of the assembly process. The multiple massive stellar nuclei observed in the centre
of Abell~3827 by \citet{Gu_2018}, for example, provide an observational analogue of the small number of massive progenitor cores that dominate BCG assembly in our hindcasting analysis.

The temporal evolution of this process can presently be tested only statistically, by comparing cluster samples across redshift rather
than by reconstructing the history of an individual system. Observational studies combining BCG masses, close companions and diffuse light already indicate that BCG and ICL growth are coupled but need not proceed at the same rate. \citet{Burke2015}, for example, find substantial growth of the ICL over $0.18<z<0.90$ while the BCG stellar mass grows much more modestly, and infer that a significant fraction of stellar material involved in interactions is deposited into the diffuse component. Similarly, \citet{Furnell2021} find strong ICL growth over $0.1<z<0.5$. Such measurements can be compared with the predicted distribution of important infall events with lookback time in Figure~\ref{fig:bcg_average}, although any quantitative comparison must apply the same observational BCG--ICL definition to the simulations.

Other quantities introduced here remain intrinsically model-dependent. Observations cannot normally determine the $R_{200}^{\rm max}$ of an individual disrupted progenitor, follow the orbital energy loss of that progenitor through time, or identify which present-day diffuse stars were associated with a particular infall event. The precise decomposition into merger-delivered and stripped populations, the BCG progenitor fraction $f_{\rm cen}$, and quantities such as $N_{50}$ are therefore most naturally regarded as simulation derived diagnostics. Their observational value lies instead in the predictions they generate: the radial variation of stellar populations, the relative kinematics of the central and diffuse components, the frequency and masses of BCG companions, and the redshift evolution of the BCG and ICL fractions. The proposed separation scheme should therefore be viewed as a physical calibration for observable proxies rather than as a classification that can be applied directly to imaging data.

Some quantitative aspects of the decomposition should consequently be regarded as model- or definition-dependent. In particular, the absolute BCG and ICL mass fractions, the precise cluster-centric crossover radius, and the detailed extent of the central component depend on the adopted BCG progenitor central radius and, to some degree, on numerical resolution and galaxy-formation physics. This is consistent with the homogeneous comparison of four hydrodynamical simulation suites by \citet{HBrown2026}, who find that absolute ICL fractions and their apparent redshift evolution are highly sensitive to the adopted BCG--ICL definition, while differences between simulations become much smaller when a common definition is imposed. Our own \textsc{GIZMO\_3k} versus \textsc{GIZMO\_7k} comparison leads to a similar distinction: the detailed scale and kinematics of the central component change modestly with resolution, whereas the qualitative separation between centrally delivered and diffusely stripped populations remains intact. We therefore regard the assembly pathways and their dynamical ordering as the robust result, rather than any particular numerical BCG--ICL boundary.

The central conclusion is therefore straightforward. The BCG--ICL division is real in terms of assembly history and dynamics, but not as a sharp structural boundary. It reflects a dynamical sorting of a common stellar population, governed by the interplay between dynamical friction and tidal stripping during hierarchical growth. Forecasting allows these components to be separated physically, even though they remain interleaved in configuration space.

\section{Conclusions}
\label{sec:conclusions}

We present a physically motivated decomposition of the central stellar component in galaxy clusters, defined by the dynamical fate of accreted systems rather than by an \emph{a posteriori} spatial cut. Using complementary hindcasting and forecasting analyses, we show that this physically defined decomposition both explains and reproduces conventional aperture-based definitions, while providing a direct interpretation in terms of hierarchical assembly.

Our main conclusions are as follows.
\begin{enumerate}
\item {\it The BCG is assembled in an oligarchic manner.} A small number of progenitors dominate its growth: in the systems studied here, half of the final BCG stellar mass is typically contributed by only 3--5 infall events. BCG assembly is therefore driven by rare, high-impact events rather than the cumulative effect of numerous minor accretions.

\item {\it The majority of accreted galaxies contribute no stellar mass to the final BCG.} Most systems survive as long-lived satellites, losing stars through tidal stripping without those stars ever becoming part of the central galaxy. Accretion into the halo is not equivalent to delivery to the centre.

\item {\it The forecasting analysis provides a physically motivated definition of the BCG:} it is the stellar component originating from the central cores of progenitor systems with low merger mass ratios whose central galaxies are delivered to the cluster centre through efficient dynamical friction prior to disruption. This selection naturally identifies a small number of dominant infall events and the radial distribution of the contributed stars does not correspond to a fixed spatial boundary.

\item {\it The ICL corresponds to the complementary population:} stars that are stripped from galaxies or groups during their passage through the cluster. They therefore populate the broader cluster potential. The BCG and ICL are thus distinguished by their assembly pathways rather than by a sharp structural boundary, and remain strongly interleaved in configuration space.

\item {\it The central BCG component} is formed from a superposition of stars formed in-situ within the cluster and similar cores brought in via a combination of major mergers and subsequent dynamical friction.

\item {\it A compact, dynamically cold, central BCG component resides at the centre of the cluster.} Within an aperture of ${\sim} 100\,\mathrm{kpc}$ the stellar mass of this component dominates over the ICL mass. An aperture selection method is highly complete and pure within such a radius. Care should be taken with dynamically disturbed central systems as they can be asymmetrical.

\end{enumerate}

The central stellar component is best understood as a superposition of two populations. One forms in-situ within the central region of the cluster; the other forms near the centres of progenitor systems and is subsequently delivered to the cluster centre through strong dynamical friction. While within a particular simulation star formation depends on the adopted subgrid star formation and feedback model, gravitational dynamics determines how these stars are redistributed, producing a continuous transition from a compact central core to an extended diffuse envelope.

In summary, the BCG--ICL distinction reflects a dynamical sorting process within hierarchical cluster assembly. Stars delivered efficiently to the central potential through a small number of major merger events form the BCG, while the remaining accreted stellar mass, stripped from infalling objects, populates the diffuse intracluster component. This framework provides a physically motivated alternative to arbitrary aperture-based definitions and explains why such definitions succeed in relaxed systems while breaking down in dynamically complex ones.
\section*{Data Availability}
Data will be shared on request to the corresponding author. Results shown in this work use data from \textsc{The Three Hundred} galaxy clusters sample. These data are available on request following the guidelines of \textsc{The Three Hundred} collaboration, at https://www.the300-project.org. 

\section*{Acknowledgements}
The authors thank the other members of the NottICL Group for helpful discussions and comments. 
We gratefully acknowledge support from the UK Science and Technology Facilities Council (STFC) under grant ST/Y509437/1. The authors acknowledge The Red Española de Supercomputaci\'on for granting computing time for running the hydrodynamical and DMO simulations of {\sc The Three Hundred} galaxy cluster project in the Marenostrum supercomputer at the Barcelona Supercomputing Center and Cibeles Supercomputers through various RES grants. The HD hydrodynamic simulations (7K) were also performed on the DIaL3 -- DiRAC Data Intensive service at the University of Leicester through the RAC15 grant: dp235, and on the Niagara supercomputer at the SciNet HPC Consortium. DIaL3 is managed by the University of Leicester Research Computing Service on behalf of the STFC DiRAC HPC Facility (www.dirac.ac.uk). The DiRAC service at Leicester was funded by BEIS (ST/K000373/1), UKRI, STFC capital funding, and STFC operations grants (ST/K0003259/1). DiRAC is part of the UKRI Digital Research Infrastructure. SciNet is funded by Innovation, Science and Economic Development Canada; the Digital Research Alliance of Canada; the Ontario Research Fund: Research Excellence; and the University of Toronto. WC gratefully thanks Comunidad de Madrid for the Atracci\'{o}n de Talento fellowship no. 2020-T1/TIC19882 and Agencia Estatal de Investigación (AEI) for the Consolidación Investigadora Grant CNS2024-154838. He further acknowledges Project PID2024-156100NB-C21 financed by MICIU/AEI /10.13039/501100011033/FEDER, EU and ERC: HORIZON-TMA-MSCA-SE for supporting the LACEGAL-III (Latin American Chinese European Galaxy Formation Network) project with grant number 101086388 and science research grants from the China Manned Space Project. 



\bibliographystyle{mnras}
\bibliography{ref} 



\appendix

\section{BCG progenitor aperture testing}
\label{App:parameter}

The principal free parameter associated with the stellar tagging in our forecasting model is the BCG progenitor central fraction, $f_{\rm cen}=r_{\rm cen}/R_{200}^{\rm max}$. Here $R_{200}^{\rm max}$ is the maximum $R_{200}$ attained by the infalling progenitor along its resolved history. Stars lying within $r_{\rm cen}$ when the progenitor crosses the host $R_{200}$, or forming within this progenitor-centred aperture after infall, are tagged as part of its central stellar component. If the progenitor subsequently reaches the central region and disappears into the main halo, these tagged stars are incorporated into the forecast BCG population. Material outside this aperture is treated as susceptible to tidal stripping and contributes to the complementary ICL population. The procedure is illustrated schematically in Figure~\ref{fig:orbit_illustration}.

\begin{figure}
\includegraphics[width=\columnwidth]{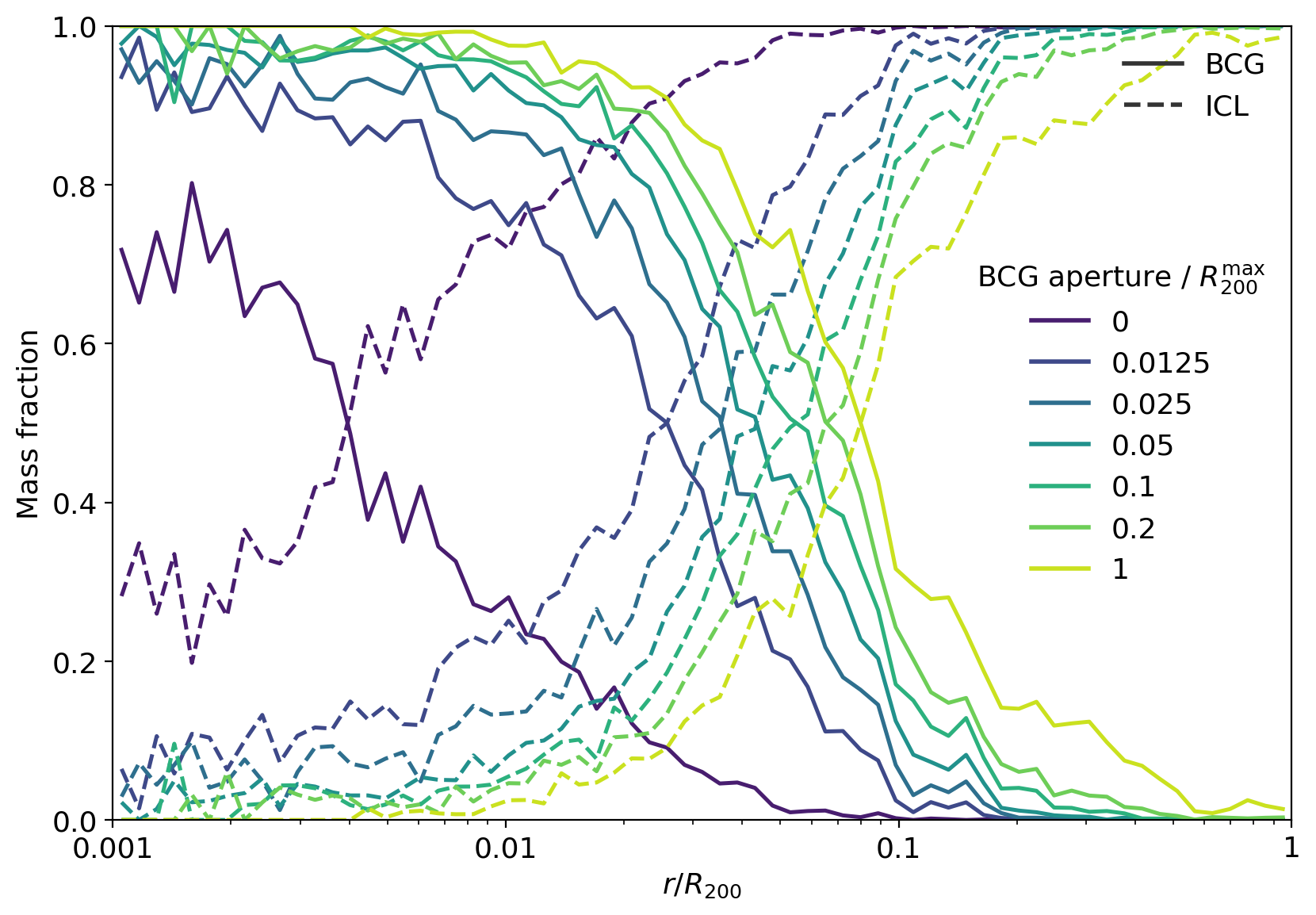}
\caption{Impact of varying the BCG progenitor central fraction $f_{\rm cen}=r_{\rm cen}/R_{200}^{\rm max}$ for cluster~2 on the radial BCG and ICL mass fractions. Solid lines show the BCG component and dashed lines the ICL component. For $f_{\rm cen}=0$, the BCG component contains only in-situ stars.}
\label{fig:partest_fractions}
\end{figure}

\begin{figure}
\includegraphics[width=\columnwidth]{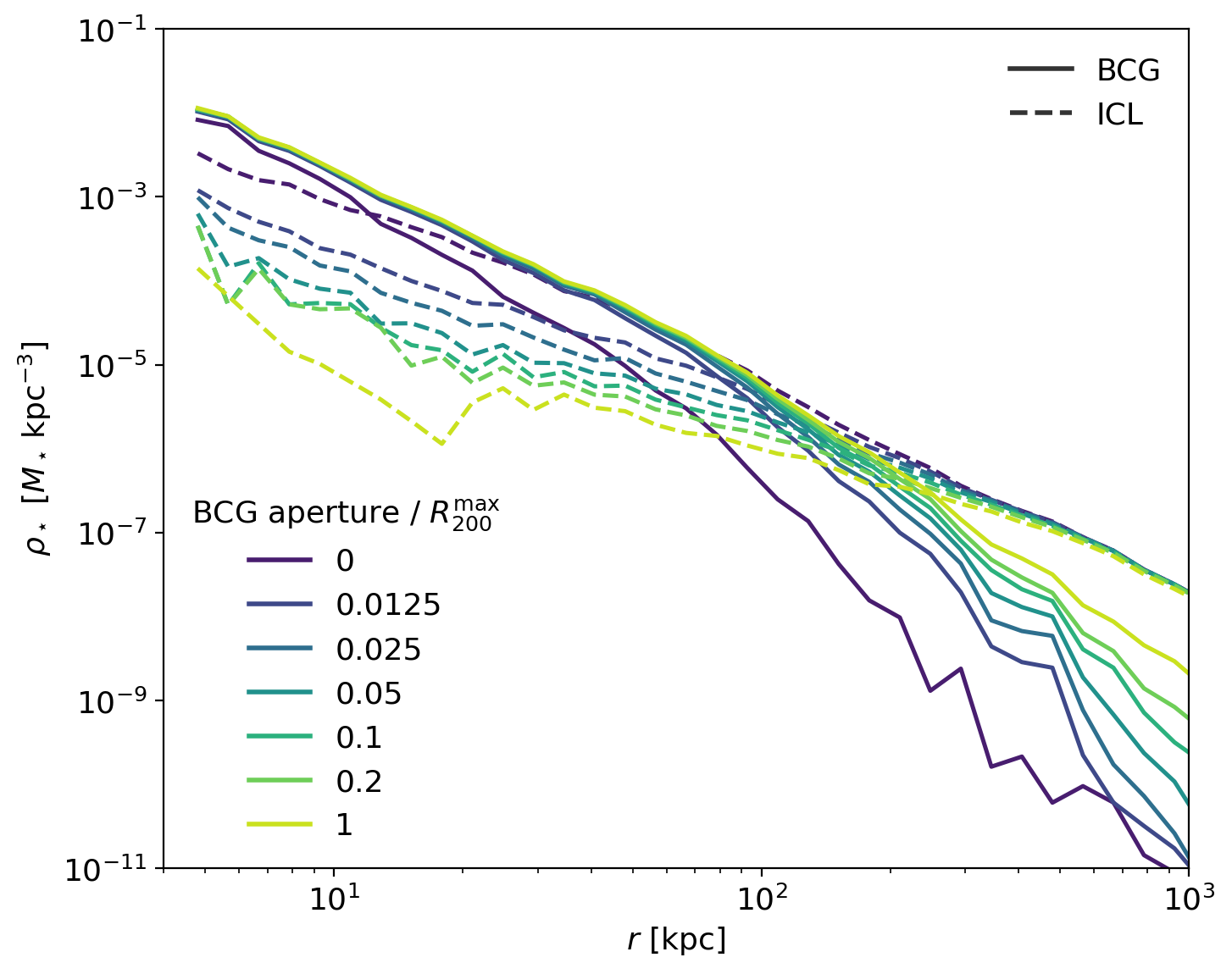}
\caption{Impact of varying the BCG progenitor central fraction $f_{\rm cen}=r_{\rm cen}/R_{200}^{\rm max}$ for cluster~2 on the density profiles of the BCG and ICL. Solid lines show the BCG component and dashed lines the ICL component. For $f_{\rm cen}=0$, the BCG component contains only in-situ stars.}
\label{fig:partest_density}
\end{figure}

Figures~\ref{fig:partest_fractions} and~\ref{fig:partest_density} show the effect of varying $f_{\rm cen}$ for cluster~2 over the full range from zero, for which only the in-situ stellar population remains in the BCG component, to unity, for which all stellar material belonging to a qualifying infalling progenitor is retained. Increasing $f_{\rm cen}$ progressively transfers more of the progenitor stellar envelope from the ICL to the BCG classification and consequently extends the forecast BCG population to larger cluster-centric radii. Conversely, for values below $f_{\rm cen}\simeq0.05$, the fraction of stars in the central region assigned to the ICL rises to approximately 10 per cent. We adopt $f_{\rm cen}=0.05$ as our fiducial value. Figure~\ref{fig:partest_density} additionally shows the corresponding density profiles; the $f_{\rm cen}=0$ case provides the radial profile of the purely in-situ central component.

We emphasise that the radii plotted in Figures~\ref{fig:partest_fractions} and~\ref{fig:partest_density} are cluster-centric radii, whereas $f_{\rm cen}$ is defined relative to the individual infalling progenitor. The transition appearing near a few per cent of the final cluster $R_{200}$ is therefore an output of the subsequent dynamical evolution rather than a radial boundary directly imposed on the final stellar distribution.

\bsp	
\label{lastpage}
\end{document}